\documentclass[3p,times,twocolumn,fleqn]{elsarticle}

\usepackage{ecrc-ac}

\usepackage{amsmath}
\usepackage{graphicx}
\usepackage{pstricks}

\usepackage{setspace}
\usepackage{wasysym}

\volume{00}

\firstpage{1}

\journalname{Nuclear Physics B Proceedings Supplement}

\runauth{P. A. Baikov et al.}

\jid{nuphbp}

\jnltitlelogo{Nuclear Physics B Proceedings Supplement}

\usepackage{amssymb}

\biboptions{sort&compress}

\usepackage[figuresright]{rotating}

\newcommand{\ice}[1]{\relax}

\newcommand{\ba}{\begin{array}}
\newcommand{\ea}{\end{array}}

\newcommand{\als}{\boldmath{\alpha_s}}

\newcommand{\g}{\gamma}

\newcommand{\bc}{\begin{center}}
\newcommand{\ec}{\end{center}}

\newcommand{\re}[1]{(\ref{#1})}

\newcommand{\ovl}[1]{\overline{#1}}

\newcommand{\sss}[1]{\scriptscriptstyle#1}

\usepackage{axodraw}
\usepackage{graphicx,longtable,amsmath,amssymb,psfrag}

\newcommand{\ed}{\end{document}}

\newcommand{\sbz}{  }

\newcommand{\nn}{\nonumber}

\ice{

}

\usepackage{axodraw}

\newcommand{\ep}{\epsilon}
\newcommand{\beq}{\begin{equation}}
\newcommand{\eeq}{\end{equation}}
\newcommand{\bea}{\begin{eqnarray}}
\newcommand{\eea}{\end{eqnarray}}

\newcommand{\bga}{\begin{gather}}

\def\ega{\end{gather}}

\newcommand{\as}{a_s}

\def\bbuildrel#1_#2^#3%
{\mathrel{\mathop{\kern 0pt#1}\limits_{#2}^{#3}}}

\catcode`\@=11
\def\slash{\mathpalette\make@slash}
\def\make@slash#1#2{\setbox\z@\hbox{$#1#2$}%
  \hbox to 0pt{\hss$#1/$\hss\kern-\wd0}\box0}
\catcode`\@=12 

\def\nnb{\nonumber}

\newcommand{\be}{\beta}

\newcommand{\promille}{%
  \relax\ifmmode\promillezeichen
        \else\leavevmode\(\mathsurround=0pt\promillezeichen\)\fi}

\newcommand{\promillezeichen}{%
  \kern-.05em%
  \raise.5ex\hbox{\the\scriptfont0 0}%
  \kern-.15em/\kern-.15em%
  \lower.25ex\hbox{\the\scriptfont0 00}}

\usepackage{fancybox}

\def\bbuildrel#1_#2^#3%
{\mathrel{\mathop{\kern 0pt#1}\limits_{#2}^{#3}}}

\allowdisplaybreaks[1]

\newcommand{\zt}{\zeta_3}
\newcommand{\zfr}{\zeta_4}
\newcommand{\zf}{\zeta_5}

\newcommand{\hs}{\hspace{-6mm}}

\begin{document}
\begin{frontmatter}



\dochead{}

\title{Massless Propagators, ${R(s)}$ and Multiloop QCD }


\author[PB]{P.~A.~Baikov}
\address[PB]{Skobeltsyn Institute of Nuclear Physics, Lomonosov Moscow State University, 
1(2), Leninskie gory, Moscow  119991, Russian Federation
}%

\author[CH]{K. G. Chetyrkin}
\address[CH]{Institut f\"ur Theoretische Teilchenphysik, Karlsruher
  Institut f\"ur  Technologie (KIT), D-76128 Karlsruhe, Germany}
\author[CH]{J. H. K\"uhn}

\begin{abstract}
 \rput(6.,10.15){\hspace{17cm}  TTP15-002  }
This is a short review of recent  developments in  calculation of
higher order corrections to various two-point corelators and  related
quantities in (massless) QCD. 


\end{abstract}


\end{frontmatter}

\section{Introduction \label{sec:1}}


Precise determinations of parameters of the Standard Model (SM) and
precise predictions for observables measured at present and future
experiments are critical in testing the SM and may hint towards
physics beyond the SM. Increasingly precise results from high energy
experiments at LEP, LHC or a future electron-positron collider have
been obtained during the past years or are expected for the coming
decade, with production and decay rates or masses of gauge or Higgs
bosons or of the top quark as characteristic examples. These are complemented
by measurements   at  low energies,  which lead to precise  values of  the strong coupling 
from $\tau$ lepton decay or the masses  of strange, charm and bottom quarks.

To extract the fundamental parameters of the theory and relate the
large number of experimental results, the knowledge of higher order
perturbative corrections is crucial. Significant advances in this
direction have been made during the past years, in particular in the
framework of the collaborative research center ``Computational
Particle Physics''  (SFB/TR-9).

\ice{
The knowledge of higher order perturbative corrections is crucial in
testing the Standard Model (SM)  and its generalizations at  high precision
 as well as in the search for new physics beyond the established
theory. 
}
Within perturbation theory quantum-theoretical amplitudes are
described by Feynman Integrals (FI's).  Improved precision, which is
required  both for strong and electroweak interactions, necessarily
leads to a significant increase of the complexity of the
calculations. This applies to the number of FI's, their increasing
complexity and, consequently, to the effort required for their
evaluation.

The complexity of a FI can  be roughly  measured by the sum of   two quantities: 
(i)   the  number    of (independent)  external momenta and
(ii)  the number of so-called ``loops'', that is the number of integrations with respect to 
internal momenta which should be performed.   In addition, the pattern of masses
of (virtual) particles appearing inside of the FI also presents  an important feature
characterizing the complexity of the integral.  

For instance, one-loop diagrams can be calculated in analytic form
for arbitrary masses and external momenta, diagrams with three or more loops,
however, only for configurations involving just one mass or energy scale 
(one-scale diagrams). Problems involving several scales, in particular those with 
pronounced scale hierarchy, can be treated approximately using asymptotic expansions 
(Hard Mass Expansion and/or Large Momentum Expansion, for example), where
each of the diagrams is expressed through a nested sum of one-scale diagrams \cite{Smirnov:2002pj}.

The present review will deal with a special class of one-scale FI's,
namely massless {\em propagators}, that is integrals depending  on only
one external momentum, $q$, and with vanishing internal masses. In what
follows we will customarily refer to massless propagator-type FI's as
{\em $p$-integrals}.

\section{Massless Propagators and Physics  \label{sec:2}   }

\subsection{RG functions and IR reduction }
The method of the renormalization group (RG)
\cite{Stueckelberg53,GellMann:1954fq,Bogolyubov:1956gh} is of vital
importance in modern quantum field theory. It is enough to recall that
the famous idea of  asymptotic freedom is based on the RG concept
of the running coupling constant.  

The RG functions ---
$\beta$-functions and various anomalous dimensions --- serve as
coefficients in the RG equations. They can be conveniently  expressed in terms of 
p-integrals (see below)
within the framework of Dimensional Regularization \cite{Ashmore:1972uj,Cicuta:1972jf,tHooft:1972fi} and 
Minimal Subtraction (MS) schemes \cite{tHooft:1973mm}. 
The naturalness and convenience of the MS-scheme for RG calculations
comes from the following statement \cite{Collins:1974da}:

\vglue 0.2cm
{\bf Theorem   1.} {\it Any UV counterterm for any FI 
integral and, consequently, any RG function in an  arbitrary minimally
renormalized model is a polynomial in momenta and masses}.
\vglue 0.1cm

\vglue 0.1cm

\noindent
This  observation was  elaborated  by
A. Vladimirov \cite{Vladimirov:1979zm}
to simplify considerably  the calculation of the  RG
functions.  The  method  was further  developed and named
Infrared Rearrangement (IRR) in \cite{Chetyrkin:1980pr}. It essentially amounts to
an appropriate transformation of the IR structure of  FI's by setting
zero some external momenta and masses (in some cases after 
the differentiation is performed with respect to the latter).
As a result  the calculation of  UV counterterms is 
reduced to  that of   $p$-integrals.  
The method of IRR was ultimately refined
and freed from unessential complications   by inventing the  so-called
$R^*$-operation \cite{Chetyrkin:1984xa}. 
The  main use of the $R^*$ -operation
is in the proof of the following statement \cite{Chetyrkin:1984xa}:
\vglue 0.2cm

\noindent
{\bf Theorem 2.} {\it Any (L+1)-loop UV counterterm for any
Feynman integral may be expressed in terms of pole and finite parts
of some appropriately  constructed L-loop  $p$-integrals.}
\vglue 0.1cm

Theorem  2  is a key tool  for  multiloop RG
calculations as it reduces  the general task of evaluation of
(L+1)-loop  UV counterterms to a well-defined and clearly posed
purely mathematical problem: the calculation of L-loop $p$-integrals.
In the following  we shall refer  to the latter as the  L-loop Problem.
A  short account of the current status of the Problem can be found in 
Section \ref{calc:methods}.


\subsection{Two-point correlators}

Within perturbation theory, every two-point correlator
\beq
\Pi^{j_1 j_2} = \int \mathrm{d}x \, e^{iqx}\,\langle 0|T\left[ j_2(x) \, j_1(0) \right] |0\rangle 
{},
\label{G:2point}
\eeq
with $j_1$ and $j_2$ being in general elementary fields or
(local) composite operators, is expressed within PT  in terms of p-integrals provided 
the momentum transfer $q$ is considered as large with respect to all
relevant masses and, thus,  the elementary field propagators contributing to $\Pi^{j_1 j_2}$ be
effectively considered as massless.

An important class of two-point correlators is represented by the case 
of $j_1$ and $j_2$ 
being quark currents of the form:
\[ 
j_1 = \ovl{\psi} \Gamma\psi, \ \  \ \ \ \  j_2= j_1^{\dagger}.
\]
In particular, the total cross-section of $e^+ e^- $ annihilation into hadrons, the
(inclusive) Higgs decay rate into hadrons, the semihadronic  decay rate of the
$\tau$ lepton coupling are all expressible
in terms of absorptive parts of the quark current correlator (\ref{G:2point}) with
$\Gamma$ chosen as $\gamma_\mu$,  $1$ and  $(1-\gamma_5)\gamma_\mu$ respectively.

Clearly, one could compute a (L+1)-loop two-point correlator by computing
the corresponding set of (L+1)-loop p-integrals. But one can  do much
better if the final aim is the absorptive part of the correlator.

Indeed, let $\Gamma$ be  a particular (L+1)-loop  Feynman diagram contributing to the
perturbative expansion of a massless correlator.  The renormalized version of the corresponding  Feynman 
integral can be generically  written as\footnote{Without essential loss of generality  we 
assume that   $\langle\Gamma\rangle(\as,Q^2\rangle$ 
is a scalar integral depending on the external momentum
$Q$ via its 
square, $Q^2 = Q_\nu Q^\nu$. In addition, we set the renormalization scale parameter $\mu=1$.} 
\beq
R \,\langle\Gamma\rangle(Q^2) = \langle\Gamma\rangle(Q^2) + \fbox{$\sum_{\gamma}Z_\gamma \langle\Gamma/\gamma\rangle(Q^2)
{} + \dots$}
\label{R}
\eeq
Here $Z_\gamma$ is the  UV Z-factor corresponding to a 1PI subgraph
$\gamma$ of $\Gamma$ and dots stand for contributions with two and
more UV subtractions. The finiteness of the left part of eq. (\ref{R})
means that the pole part in $\ep = (4-D)/2$ of $\langle\Gamma\rangle(Q^2)$ is
completely fixed by poles in $\ep$ which appear in UV subtractions (the
boxed terms in (\ref{R})). On the other hand, the UV subtractions
could, obviously, contain $L'$-loop Z-factors with $L' \le L+1$ and
the reduced p-integrals like $\langle \Gamma/\gamma\rangle(Q^2)$ with the loop
number {\em not exceeding} L! Applying Theorem 2 we arrive  at the 
conclusion that the pole part of $\langle\Gamma\rangle(Q^2)$ (and, consequently,
its absorptive part) is completely expressed via L-loop p-integrals
{\em only}.

We want also to stress  that by  high-energy limit we understand not
only the case when all  masses can be neglected but also  the   possibility 
to take into account mass effects by exploiting a small mass  expansion.
As  a suitable example one could mention the calculation
of the power suppressed (of order $m_q^2/s$, $m_q^4/s^2$ and so on) corrections for
the correlators of (axial)vector quark currents in higher orders of pQCD 
\cite{ChetKuhn90,Chetyrkin:1997qi,Chetyrkin:2000zk,Baikov:2004ku,Baikov:2009uw}.

\subsection{OPE and DIS}
The   theoretically cleanest  description 
of  (inclusive)   Deep  Inelastic Scattering (DIS) can be achieved within 
Operator Product Expansion (OPE) of two composite operators 
(for  a recent  review see, e.g \cite{Blumlein:2012bf}).
Here the main objects to compute are the so-called Coefficient Functions (CF)
which can  be always computed via 
p-integrals with the help of the so-called method of projectors
\cite{Gorishnii:1983su,Gorishnii:1986gn}. It is important to stress that
within the method of projectors one needs no IRR: L-loop corrections to  a CF
can be expressed  {\em directly} in terms of L-loop p-integrals.   

A good example of an early multiloop OPE calculation is the one of the
$\als^3$ corrections to the Bjorken sum rule for polarized
electroproduction and to the Gross-Llewellyn Smith sum rule
\cite{Larin:1991tj}. We will discuss  later our calculations of the
next, order $\als^4$, contributions to the  Bjorken sum rule.

\section{Calculational Methods \label{calc:methods}}

A significant number of higher order calculations are usually
performed according to the following ``standard'' scenario. First, the
Feynman amplitudes are reduced to a limited set of so-called {\em
master integrals} (MI's). This step is based on recursion algorithms
obtained by using Integration-by-Parts (IBP) identities (see, e.g. 
recent books and reviews \cite{Smirnov:2006ry2,Smirnov:2012gma2,Grozin:2007zz,Grozin:2011mt}
 and references therein).

An important feature of the standard scenario is that the resulting
set of master integrals should be computed only once and forever due
to the well-established property of universality: for every
given class of Feynman amplitudes characterized by the number of loops
and the pattern of external momenta and masses 
the corresponding set of master integrals is universal in the following
sense: every (even extremely complicated)  amplitude from the class can be
expressed in terms of  one and the same (finite! \cite{Smirnov:2010hn,Lee:2013hzt}) set of master
integrals .

Thus, the task of evaluation of p-integrals at L-loops (L-loop
Problem) is naturally decomposed in two:
(A) reduction of
a  generic L-loop p-integral  to masters   and
(B) evaluation of the latter.   

Both A and B Problems  were solved at two- and three-loop level long ago \cite{Chetyrkin:1980pr,Chetyrkin:1981qh}.
The four-loop problem has been under  active investigation in our group since  the beginning of the
current century. We will describe  the current status in the next two Sections.

\subsection{A: Reduction via $1/D$ expansion}
The standard (Laporta) \cite{Laporta:1996rh,Laporta:2000dc}  
approach to solve IBP relations implies a step-by-step linear
reduction of more
complicated integrals to less complicated and finally to irreducible
(master) integrals.
Unfortunately this conceptually simple method could not be used in our
case because of the 
extremely large amount ($10^7-10^8$) of 4-loop integrals appearing after
IRR for a typical
5-loop problem. As the result we used a more sophisticated and laborious, but
less demanding for  computer resources method based on large $D$
expansion \cite{Baikov:2005nv} of the formal solutions of the IBP relations
\cite{Baikov:1996rk}.

So assume that we need to perform the reduction,  that is to calculate
the coefficients in front of master integrals. The coefficients depend
on indices of the original integral and fulfill the IBP identifies
according to these indices. If we construct some ``convenient''
solutions of the identities, then the coefficients we need can be obtained as
their linear combinations with proper boundary conditions.

As shown in \cite{Baikov:1996rk} such ``convenient'' solutions can be
constructed in the form of the integrals of polynomial raised to some degree
(linear in dimension $D$). Unfortunately, although these  integrals are simpler
than the original FI, they are still too complicated for direct evaluation. From
the other side, we are interested  in linear combinations of these integrals
which are rational over $D$ (because in principle they  can be calculated by
the step-by-step reduction). So one can expand these integrals in the $1/D$ limit
(resulting in integrals of Gaussian type \cite{Baikov:2005nv}),
calculate sufficiently many terms and finally reconstruct the exact  $D$
dependence.

Note, that we need not calculate each of the  $10^7-10^8$ integrals involved in
the  specific problem individually (as it is necessary in the standard
reduction). Instead we can calculate $1/D$ expansion coefficients of the
total expression we are interested  in, thus saving a lot of computer
resources.

The construction of the large-$D$ limit requires in general huge 
storage resources, which naturally constrains the structure of the 
input p-integrals: they should better not contain any  extra parameters like
color coefficients, $n_f$, the number of light quark  flavours contributing to  internal fermion
loops, and so on. As a result we are forced to use a ``slice'' approach:
that is to set all color   coefficients  to their  numerical  values and 
to fix $n_f$ to some integer. Combining together different slices one can
always  reconstruct the full $n_f$-dependent structure (and, if necessary, even all colour coefficients)

\subsection{B: Master p-integrals and their evaluation}
\begin{center}
\begin{figure}
\hspace{3mm}\includegraphics[width=70mm]{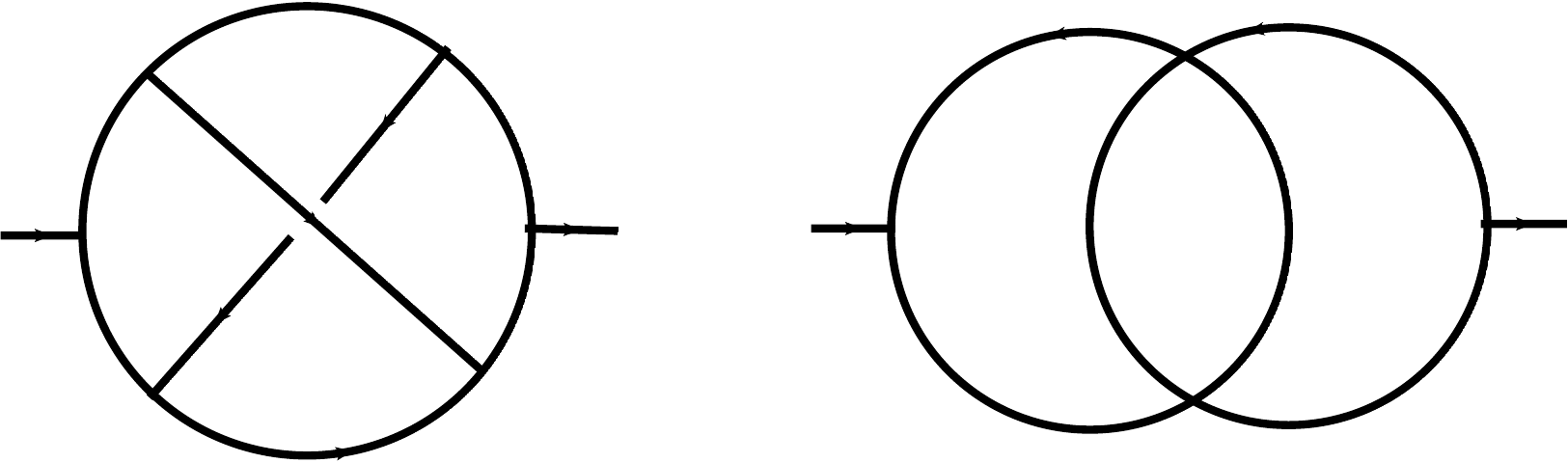}
\caption{Non-trivial three-loop master p-integrals. }
\label{3:loop:masters}
\end{figure}
\end{center}

At one- at two-loop levels the master p-integrals are trivial as they can  be easily performed   
in terms of $\Gamma$-functions for a generic value of the running space-time dimension  $D$ (see, e.g. 
\cite{Grozin:2003ak}. At three loops there exist only two non-trivial master p-integrals (see Fig. 1);
their values are  known since long \cite{Chetyrkin:1980pr}. 

The significantly more complicated problem of identifying and computing
{\em all} 28 four-loop master p-integrals has been solved only
recently. We refer the interested reader to the original publications
\cite{Baikov:2005nv,Baikov:2010hf,Smirnov:2010hd,Lee:2011jt}.


\subsection{Computer Algebra \& FORM}

Higher order calculations dealing with thousands of diagrams already at 3-loop level 
require   heavy use  of computer algebra tools. We are using QGRAF \cite{QGRAF}
for automatic  diagram generation   as well as a collection of Mathematica  and PERL 
scripts to automatically assign topologies and prepare input files for FORM.

The workhorse for all the  complicated calculations 
discussed  in the current  paper 
 is the computer algebra program {\tt
 FORM}~\cite{Vermaseren:2000nd} and its parallel versions {\tt
 ParFORM}~\cite{Tentyukov:2004hz} and {\tt TFORM}~\cite{Tentyukov:2007mu}.
The program offers excellent possibilities for dealing with gigantic
data streams generated during the reduction procedure. 
The internal specifications 
allow FORM to deal with expression which are much larger than
the available memory (RAM). The only restriction for the size of an
expression is the disk space which nowadays 
is rather cheap. As a consequence, the complexity of a problem
solvable by FORM is practically restricted only by time.

The FORM program, BAICER,  intended   for the reduction of complicated four-loop
p-integrals implements the algorithms described in the previous Section.  With
increasing experience and using heuristic criteria about the pole
structure of the coefficient functions in front of master integrals,
BAICER has developed into an efficient tool which allows to calculate
complicated four-loop massless propagator integrals, including their
finite part. It runs routinely on ParFORM and TFORM using 8 to 16
cores with a speed-up between 6 and 12.

\section{Scalar  Correlator \& $\Gamma(H \to \ovl{q}q)$ }
The decay width of the Higgs boson into  a
pair of quarks can be written in the form
\beq
\Gamma(H \to \bar{f}f )
=\frac{G_F\,M_H}{4\sqrt{2}\pi}\,
m_f^2(\mu) \,  R^S (s = M_H^2,\mu)
\label{H2ff}
\eeq
where $\mu$ is the
normalization scale and
\beq
 R^S(s)
 = 
\mbox{\rm Im} \,  {\Pi^{SS}}(-s-i\epsilon)/{(2\pi\,  s)} 
\label{scalar:densiry}
\eeq
 is the spectral density of the scalar correlator
\beq
\label{Pi:S}
{\Pi^{SS}} (Q^2)
= (4\pi)^2 i\int dx e^{iqx}\langle 0|\;T[\;J^{\rm S}_f(x)
J^{\rm S}_{f}(0)\,]\;|0\rangle
{}. 
\eeq
Here $Q^2 = - q^2$ and $J^{\rm S}_f=\ovl{\Psi}_f\Psi_f$ is the scalar
current for quarks with flavour $f$ and mass $m_f$, coupled to the
scalar Higgs boson. The ${\cal O}(\alpha_s^4)$ result for $R^S$ is
known analytically since long \cite{Baikov:2005rw, Chetyrkin:2005kn}
(early results for orders $\alpha_s^2$ and $\alpha_s^3$ can be found in
\cite{Gorishnii:1990zu} and \cite{Chetyrkin:1996sr} respectively).
For brevity we put below the final result in numerical form:
\begin{gather}
\widetilde{R}(s, \mu^2 =s) = 
1
+
5.6667  a_s
{+} 
\left[
35.94
-1.359  \, n_f
\right]
a_s^2
\nonumber
\\
{+}  a_s^3
\left[
164.14
-25.77  \, n_f
+0.259  \, n_f^2
\right]
\label{RSnum}
\\
{+} \,a_s^4
\left[
39.34
-220.9  \, n_f
+9.685  \, n_f^2
-0.0205  \, n_f^3
\right]
{}.
\nonumber
\end{gather}
We will discuss the application of \re{RSnum} for the dominant b-quark decay mode of the Higgs
boson later in Section~\ref{gamma:m:pheno}.
\section{Vector Correlator \& $R(s)$}
The ratio 
\[R(s)\equiv \sigma(e^+e^-\to {\rm hadrons}) / \sigma(e^+e^-\to\mu^+\mu^-)
\] 
is 
expressed through the absorptive part of the 
vector correlator 
\begin{align} 
\nnb
\Pi_{\mu\nu}(q) & =
 i \int {\rm d} x e^{iqx}
\langle 0|T[ \;
\;j_{\mu}^{\rm em}(x)j_{\nu}^{\rm em}(0)\;]|0 \rangle
=
\\
&=
\displaystyle
(-g_{\mu\nu}q^2  + q_{\mu}q_{\nu} )\Pi(-q^2)
{}\, ,
\label{PI}
\end{align}
with the hadronic EM current
$
j^{\rm em}_{\mu}=\sum_{{f}} Q_{{f}}
\overline{\psi}_{{f}} \gamma_{\mu} \psi_f
$, and $Q_f$ being the EM charge of the quark $f$.
The optical theorem
relates  the inclusive cross-section
and thus the function $R(s)$
to the discontinuity of $\Pi$
in the complex plane
\begin{equation} 
R(s) =  \displaystyle
 12 \,\pi \,{\rm Im}\, \Pi( - s -i\delta)
\label{discontinuity}
{}\, .
\end{equation}

\begin{figure}
\begin{center}
\includegraphics[width=75mm]{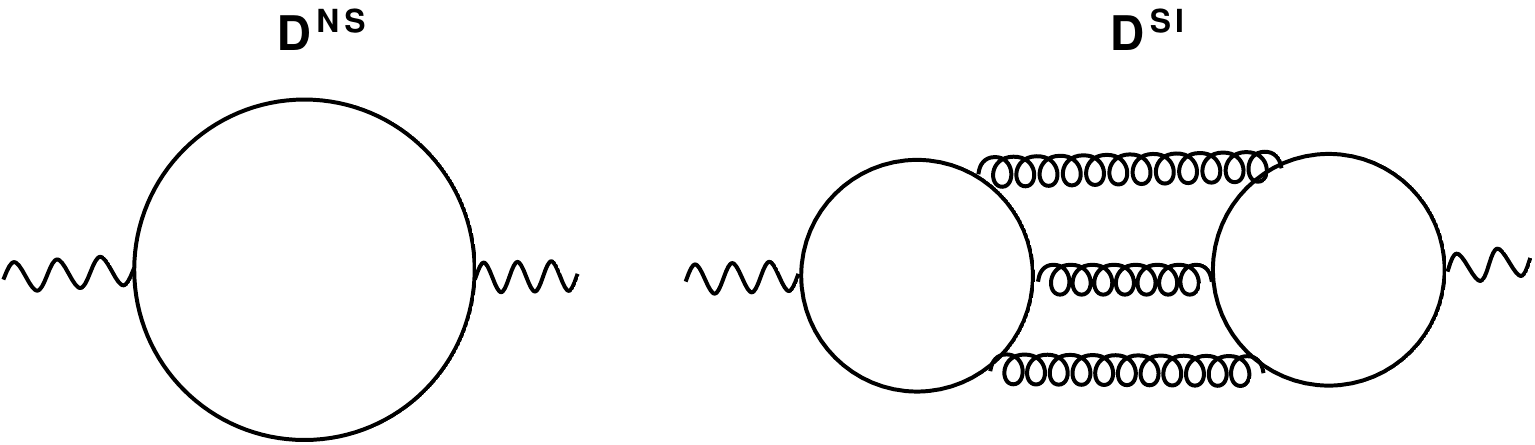}
\end{center}
\caption{
Lowest order non-singlet (a) and singlet (b) diagrams contributing to the polarization operator.
}
\label{figure:1}
\end{figure}

For the vector correlator the terms of order $a_s^2$ and $a_s^3$ are  known 
since  long \cite{Chetyrkin:1979bj,Gorishnii:1991vf}.
The next,  $a_s^4$ order   has been under  investigation  for more then 10 years 
\cite{Baikov:2001aa,Baikov:2002uw,Baikov:2002va,Baikov:2003gu,Baikov:2004ku,Baikov:2005sw,Baikov:2006nb,Baikov:2008jh,Baikov:2009uw,Baikov:2012zm,Baikov:2010iw,Baikov:2012er,Baikov:2012zn}
in  A1 group. By now it is  known in  a complete form for   a generic colour  group $G$ \cite{Baikov:2012er,Baikov:2010iw}. We put below
only physically relevant result for $G=SU(3)$:
\begin{align}
& \hspace{-6mm} R(s) =~ 3\sum_f Q_f^2 \bigg\{
1 + a_s + 
\label{Rqcd}
\\[1mm]
& \hs+ a_s^2 \Big( \tfrac{365}{24} - 11\,\zeta_3 - \tfrac{11}{12}\,n_f + \tfrac{2}{3}\,\zeta_3 \,n_f \Big)
\nonumber\\
& \hs~+ a_s^3\,\Big[n_f^2\,\Big( \tfrac{151}{162} - \tfrac{1}{108}  \pi^2 - \tfrac{19}{27}\,\zeta_{3}\Big)
\nonumber \\[1mm]
& \hs\qquad  + n_f\,\Big( - \tfrac{7847}{216} + \tfrac{11}{36}\,\pi^2 + \tfrac{262}{9}\,\zeta_{3} - \tfrac{25}{9}\,\zeta_{5}\Big)
\nonumber \\[1mm]
& \hs\qquad + \tfrac{87029}{288} - \tfrac{121}{48}\,\pi^2 - \tfrac{1103}{4}\,\zeta_{3} + \tfrac{275}{6}\,\zeta_{5} \Big]
\nonumber \\[1mm]
& \hs~+ a_s^4\,\Big[ n_f^3\,\Big( - \tfrac{6131}{5832} + \tfrac{11}{432}\,\pi^2 + \tfrac{203}{324}\,\zeta_{3} - \tfrac{1}{54}\,\pi^2\,\zeta_{3}
  + \tfrac{5}{18}\,\zeta_{5}\Big)
\nonumber\\[1mm]
& \hs\qquad + n_f^2\,\Big( \tfrac{1045381}{15552} - \tfrac{593}{432}\,\pi^2 - \tfrac{40655}{864}\,\zeta_{3}
\nonumber\\[1mm]
& \hs\qquad\qquad\qquad + \tfrac{11}{12}\,\pi^2\,\zeta_{3}  + \tfrac{5}{6}\,\zeta_3^2 - \tfrac{260}{27}\,\zeta_{5}\Big)
\nonumber\\[1mm]
& \hs\qquad + n_f\,\Big( - \tfrac{13044007}{10368} + \tfrac{2263}{96}\,\pi^2 + \tfrac{12205}{12}\,\zeta_{3} - \tfrac{121}{8}\,\pi^2\,\zeta_{3}
\nonumber\\[1mm]
& \hs\qquad\qquad\qquad - 55\,\zeta_3^2 + \tfrac{29675}{432}\,\zeta_{5} + \tfrac{665}{72}\,\zeta_{7}\Big)
\nonumber\\[1mm]
& \hs\qquad + \tfrac{144939499}{20736} - \tfrac{49775}{384}\,\pi^2 - \tfrac{5693495}{864}\,\zeta_{3} + \tfrac{1331}{16}\,\pi^2\,\zeta_{3}
\nonumber\\[1mm]
& \hs\qquad\qquad\qquad  +\tfrac{5445}{8}\,\zeta_3^2 + \tfrac{65945}{288}\,\zeta_{5} - \tfrac{7315}{48}\,\zeta_{7}\Big]
\bigg\}
\nonumber\\[1mm]
& \hs~ + \bigg(\sum_f Q_f \bigg)^2 \bigg\{ a_s^3 \Big( \tfrac{55}{72} - \tfrac{5}{3}\,\zeta_3 \Big)
\nnb
\\[-2mm]
& \hs\qquad\qquad + a_s^4\,\Big[ n_f\,\Big( - \tfrac{745}{432} + \tfrac{65}{24}\,\zeta_{3} + \tfrac{5}{6}\,\zeta_3^2 - \tfrac{25}{12}\,\zeta_{5}\Big)
\nonumber\\
& \hs\qquad\qquad\qquad + \Big( \tfrac{5795}{192} - \tfrac{8245}{144}\,\zeta_{3} - \tfrac{55}{4}\,\zeta_3^2
  + \tfrac{2825}{72}\,\zeta_{5}\Big)\Big]\bigg\}~,
\nonumber
\end{align}
where $a_s \equiv \alpha_s/\pi$ and we have set the normalization
scale $\mu^2= s$;  the results for generic values of $\mu$ can be
easily recovered with standard RG techniques.  Note that the two specific
quark charge structures in (\ref{Rqcd}) correspond to the  so-called
 non-singlet (numerically dominant) and  the singlet contributions (see Fig.~\ref{figure:1}) to  the 
vector correlator (\ref{PI}).
Numerically, 
\begin{align}
& \hs  R(s) =~ 3\sum_f Q_f^2\,\bigg\{
1 + a_s + a_s^2\,\Big( 1.986 - 0.1153 \,n_f \Big)
\nonumber\\[1mm]
& \hs \qquad+ a_s^3\,\Big( -6.637 - 1.200 \,n_f - 0.00518 \,n_f^2 \Big)
\label{Rfull_numeric}
\\[1mm]
& \hs \qquad +a_s^4\,\Big( -156.608 + 18.7748\,n_f  - 0.797434 \,n_f^2
\nonumber\\
& \hs \qquad\qquad + 0.0215161 \,n_f^3 \Big)\bigg\}
\nonumber
\\
& \hs ~- \bigg(\sum_f Q_f\bigg)^2 \Big( 1.2395\, a_s^3   + \Big(17.8277 - 0.57489 \,n_f \Big)\, a_s^4 \Big)~.
\nonumber
\end{align}
Specifically, for the  particular values of $n_f = 3,4 $ and 5 one  obtains (for the terms of order $\alpha_s^3$ and  $\alpha_s^4$ we have explicitly decomposed the coefficient into  non-singlet and singlet contributions):
\ice{
Out[23]//InputForm= 
1. + as + 1.639821204896986*as^2 - 10.283942943737344*as^3 - 
 106.87979007077956*as^4
}
\begin{gather}
R^{n_f=3}(s)  =  2\, \bigg[ 1 + a_s + 1.6398 a_s^2 - 10.2839 a_s^3
\\
\nnb
  \hspace{5cm}    -  106.8798 a_s^4 \bigg]
\end{gather}
\vspace{-16mm}
\begin{align}
\nonumber \\
&R^{n_f=4}(s)  =~ \frac{10}{3} \bigg[ 1 + a_s + 1.5245 a_s^2
\nonumber \\
&~+ a_s^3\, \Big( -11.686 =-11.52 - 0.16527^{\rm SI}\Big)
\nonumber \\
&~+ a_s^4\, \Big(  - 94.961 = -92.891 - 2.0703^{\rm SI}\Big) \bigg]~,
\\
& R^{n_f=5}(s)  =~ \frac{11}{3} \bigg[ 1 + a_s + 1.40902 a_s^2
\nonumber \\
&~+ a_s^3\, \Big( - 12.80=-12.767 - 0.037562^{\rm SI}\Big)
\nonumber \\
&~+ a_s^4\, \Big(  - 80.434 = -79.981 - 0.4531^{\rm SI}\Big) 
\bigg]
{}.
\end{align}
Note that for $n_f=3$ the singlet contributions vanish in every order in $\alpha_s$ as the corresponding
global coefficient $(\sum_f Q_f)^2$ happens to be zero. 
Implications of this result for the determination of $\alpha_s$ in 
electron-positron annihilation and in $Z$-boson decays are discussed in
\cite{Baikov:2008jh,Baikov:2012er}.

As a by-product of the calculation of $R(s)$ the authors of \cite{Baikov:2012zm}
have obtained the five-loop $\beta$-function in pure QED, that is a
theory with $n_f$ single-charged fermions minimally coupled to the
photon field.  The result reads (the four-loop result is known since long from 
\cite{Gorishnii:1991kd})
\bea
&{}& \beta^{\sss QED}(A) = 
\,n_f 
\left[
\frac{4\ A^2}{3}  
\right]
{+}  
4\,  n_f A^3
-  A^4
\left[
2  \,n_f 
+\frac{44}{9}  \, n_f^2
\right]
\nonumber\\
&{+}& \hspace{-2mm} A^5
\left[
-46  \,n_f 
+\frac{760}{27}  \, n_f^2
-\frac{832}{9}  \,\zeta_{3} \, n_f^2
-\frac{1232}{243}  \, n_f^3
\right]
\nonumber\\
&{+}& \hspace{-2mm} A^6\Biggl( 
\, n_f^3
\left[
-\frac{21758}{81} 
+\frac{16000}{27}  \sbz \zeta_{3}
-\frac{416}{3}  \sbz \zeta_{4}
-\frac{1280}{3}  \sbz \zeta_{5}
\right]
\nonumber\\
&& 
\hspace{2mm}
{+} \, n_f^2
\left[
-\frac{7462}{9} 
-992  \sbz \zeta_{3}
+2720  \sbz \zeta_{5}
\right]
\nonumber\\
&{}& 
\hspace{2mm}
+\,n_f 
\left[
\frac{4157}{6} 
+128  \sbz \zeta_{3}
\right]
{+} \, n_f^4
\left[
\frac{856}{243} 
+\frac{128}{27}  \sbz \zeta_{3}
\right]
\,
\Biggr)
{}.
\eea 
Here  the QED coupling constant 
\[
A(\mu) = \alpha(\mu)/(4\,\pi)=e(\mu)^2/(16\,\pi^2)
{}.
\]

\section{QCD RG-functions}

Our starting point is the QCD Lagrangian with $n_f$ quark  flavours written in terms of 
renormalized fields, coupling constant $g$ and quark masses $m_f$:
\begin{align} 
{\cal L}_0  &=- \frac{1}{4} Z_3\, ( \partial _{\mu}A_{\nu} 
-  \partial _{\nu}A_{\mu})^2
\nnb
\\
&- \frac{1}{2}g\, Z_1^{3g} \, ( \partial _{\mu}A^a_{\nu} 
-  \partial _{\nu}A^a_{\mu})
\,  ( A_{\mu} \times A_{\nu})^a
\nnb\\ 
&- \frac{1}{4} g^2\, Z^{4g}_1\, ( A_\mu \times A_\nu)^2
-  \frac{1}{2  \xi_L }  ( \partial _\nu A_\mu)^2
\label{QCDlag}
\\
&+ Z^c_3\, \partial_\nu \bar c  \, (\partial_\nu c )
+ g\, Z_1^{ccg} \, \partial^\mu \bar c \, (A \times   c  )
\nnb\\
&+\sum_{f=1}^{n_f} 
\bar \psi^f ( \mathrm{i} Z_2 \, \slash{ \partial } 
 + g Z^{\psi\psi g}_1  \slash{A} - Z_{\psi\psi}\, m_f )\, \psi^f
\nnb
{},
\end{align}
with 
\[
(A \times B) = f^{abc}A^b B^c, \ \slash{ \partial} = \gamma^{\mu} \frac{\partial}{\partial x_{\mu}}
\]
and 
with bare  gluon, quark and ghost fields related to the renormalized ones as follows:
\begin{equation} 
A^{a \mu}_0 = \sqrt{Z_3}\ A^{a \mu},
\ \
\psi^f_0  = \sqrt{Z_2}\ \psi^f_0,
\ \
c^{a}_0   = \sqrt{Z_3^{c}}\ c^{a} 
{}.
 \end{equation}

The   vertex Renormalization Constants (RCs)
 \begin{equation} 
Z^V_1, \ \ \ V\in \{\mathrm{3g,\ 4g,\  c  c  g  , \ \psi  \psi g}\}
{}
 \end{equation} 
renormalize 3-gluon, 4-gluon, ghost-ghost-gluon, 
quark-quark-gluon vertex functions respectively. 
The  Slavnov-Taylor identities allow one to express  all   vertex
RCs in terms of wave function RCs and  an  independent  charge 
RC,  $Z_g = \frac{ \displaystyle  g_0 }{ \displaystyle  g}$:
 \begin{eqnarray} 
Z_\xi &=& Z_3, 
\label{WI:xi}
\\
Z_g &=& \sqrt{Z_1^{4g}} \,  (Z_3)^{-1}, \ \ 
\label{WI:4g}
\\
Z_g &=& Z_1^{3g} (Z_3)^{-3/2}, \ \ 
\label{WI:3g}
\\
Z_g &=& Z_1^{ccg} (Z_3)^{-1/2} (Z_3^c)^{-1}, \ \ 
\label{WI:ccg}
\\
Z_g &=& Z_1^{\psi\psi g} (Z_3)^{-1/2} (Z_2)^{-1}
\label{WI:qqg}
{}.
 \end{eqnarray}

Within the commonly accepted $ \overline{\mbox{MS}} $ scheme RCs are  independent of 
dimensional parameters (masses and momenta) and can be represented
as follows
 \beq
Z(h) = 1 + \sum_{i,j}^{1 \le j \le i} Z_{ij}  \frac{h^i}{ \epsilon^j}
\label{}
{},
 \eeq
where $h = g^2/(16 \pi^2) = \alpha_s/(4\pi)$ and the parameter $ \epsilon $ is related to the continuous   
space time dimension $D$  via $D= 4 - 2 \epsilon $.
Given a RC $Z(h)$,  the  corresponding anomalous dimension is defined as
 \begin{equation} 
\gamma(h) = -\mu^2\frac{\mathrm{d} \log Z(h)}{\mathrm{d} \mu^2}
= \sum_{n=1}^\infty  Z_{n,1}\, n\, h^n
= -\sum_{n=0}^\infty (\gamma)_n \, h^{n+1} 
\label{anom:dim:generic}
{}.
\end{equation}
The anomalous dimension of the quark-gluon coupling constant  $h$ 
is conventionally  referred to as ``QCD $\beta$-function'';  equation
\re{WI:ccg}  leads to the calculationally simplest presentation of the function:
\beq
\beta(h)
 = 
 2\gamma_1^{ccg} - 2 \  \gamma_3^{c} -  \, \gamma_3 
{}.
\eeq

The  quark mass anomalous dimension,  $\gamma_m$, governs the evolution of  
the quark mass, viz.
\begin{equation}
\mu^2\frac{d}{d\mu^2} {m}|{{}_{{h^0},
 h^0 }}
 = {m} \gamma_m(h) \equiv
-{m}\sum_{i\geq0}\gamma_{{i}}
\,
h^{i+1}
{}.
\label{anom-mass-def}
\end{equation}
To calculate  $\gamma_m$ one needs to 
find the so-called quark mass renormalization constant, 
$Z_{m}$, which is defined as the ratio of the bare and renormalized
quark masses, viz.  
\beq
Z_m = \frac{m_f^0}{m_f} = \frac{Z_{\psi\psi}}{Z_2}
\label{Zm}
{}.
\eeq
The final formula for $\g_m$ follows from  the QCD Lagrangian \re{QCDlag} and reads
\beq
\g_m =  \g_{\psi\psi}  - \g_2
{}.
\eeq

Thus, to compute the QCD $\beta$-function and the quark mass anomalous
dimension at five loops \footnote{Up to and including four loop level they are known since
long \cite{Gross:1973id,Politzer:1973fx,Caswell:1974gg,Jones:1974mm,Egorian:1978zx,Tarasov:1980au,%
Larin:1993tp,vanRitbergen:1997va,Czakon:2004bu,Tarrach:1980up,Tarasov:1982gk,Larin:1993tq,Chetyrkin:1997dh,Vermaseren:1997fq}.}
one should
evaluate five separate anomalous dimensions, viz.
\[ \gamma_1^{ccg}, \  \gamma_3^{c},   \, \gamma_3,\  g_{\psi\psi}, \g_2 .\]
By now we have computed all of them except (most difficult)
$\gamma_3$.  The results are presented in the four next
subsections\footnote{Note that in all calculations we have used the
simplest --- Feynman ---  gauge fixing condition.  The physically relevant
$\g_m$ and $\beta$  functions do not depend on  gauge.}.

\subsection{Five-loop running of the  ghost  field}\

\begin{gather}
\g_3^c = -\sum_{i=0}^{\infty} \Big( \g_3^c \Big)_i\, h^{i+1}
{},
\\
(\g_3^c)_0 =-\frac{3}{2}, 
\\
(\g_3^c)_1 = -\frac{147}{8}
+\frac{5}{4}\ n_f 
{},
\\
(\g_3^c)_2 = 
-229 
-\frac{81}{4}  \sbz \zeta_{3}
{+} \, n_f 
\Bigg(
\frac{1085}{48} 
+\frac{33}{2}  \sbz \zeta_{3}
\Bigg)
+ 
 \frac{35}{36 }\, n_f^2
{},
\label{gcc3l}
\\
(\g_3^c)_3 = 
-\frac{1016843}{192} 
-\frac{129825}{32}  \sbz \zeta_{3}
\\
\nnb
\hspace{29mm}
+\frac{9963}{32}  \sbz \zeta_{4}
+\frac{78705}{16}  \sbz \zeta_{5}
\\
\nnb
\hspace{10mm}{+}  n_f\, 
\Bigg(
\frac{198229}{192} 
+\frac{48461}{48}  \sbz \zeta_{3}
\\
\nnb
\hspace{29mm}
-\frac{4797}{16}  \sbz \zeta_{4}
-\frac{3355}{4}  \sbz \zeta_{5}
\Bigg)
\\
\nnb
{+}  n_f^2\,
\Bigg(
-\frac{3385}{144} 
-\frac{49}{2}  \sbz \zeta_{3}
+\frac{33}{2}  \sbz \zeta_{4}
\Bigg)
{+} n_f^3\,
\Bigg(
\frac{83}{108} 
-\frac{4}{3}  \sbz \zeta_{3}
\Bigg)
{},
\end{gather}
\begin{gather}
(\g_3^c)_4 = 
-\frac{193301287}{2048} 
-\frac{19562145}{128}  \sbz \zeta_{3}
\\
\nnb
\hspace{11mm}
-\frac{2060829}{128}  \,\zeta_3^2
+\frac{1101573}{16}  \sbz \zeta_{4}
+\frac{66632427}{128}  \sbz \zeta_{5}
\\
\nnb
\hspace{11mm}
-\frac{36327825}{256}  \,\zeta_{6}
-\frac{140900823}{512}  \,\zeta_{7}
\\
\nnb
{+}  n_f \,
\Bigg(
\frac{633704171}{27648} 
+\frac{5166473}{144}  \sbz \zeta_{3}
+\frac{233519}{64}  \,\zeta_3^2
\\
\nnb
\hspace{15mm}
-\frac{764949}{32}  \sbz \zeta_{4}
-\frac{32902291}{384}  \sbz \zeta_{5}
+\frac{4123825}{128}  \,\zeta_{6}
\\
\nnb
\hspace{50mm}
+\frac{14425075}{384}  \,\zeta_{7}
\Bigg)
\\
\nnb
{+} \, n_f^2\,
\Bigg(
-\frac{1326547}{3456} 
-\frac{1739167}{864}  \sbz \zeta_{3}
-\frac{2659}{6}  \,\zeta_3^2
\\
\phantom{+ \, n_f^2}
\hspace{2.5mm}+\frac{13485}{8}  \sbz \zeta_{4}
+\frac{8074}{9}  \sbz \zeta_{5}
-\frac{16775}{12}  \,\zeta_{6}
\Bigg)
\nonumber\\
\nnb
{+} \, n_f^3\,
\Bigg(
-\frac{342895}{7776} 
-\frac{1211}{18}  \sbz \zeta_{3}
-\frac{5}{2}  \sbz \zeta_{4}
+\frac{284}{3}  \sbz \zeta_{5}
\Bigg)
\\
\nnb
\hspace{26mm}
{+}  n_f^4\,
\Bigg(
\frac{65}{108} 
+\frac{20}{27}  \sbz \zeta_{3}
-\frac{4}{3}  \sbz \zeta_{4}
\Bigg)
{},
\end{gather}

\subsection{Five-loop running of the  ghost-ghost-gluon  vertex}
\begin{gather}
\g_1^{ccg} = -\sum_{i=0}^{\infty} \Big( \g_1^{ccg} \Big)_i\, h^{i+1}
{},
\\
(\g_1^{ccg})_0 =\frac{3}{2}, 
\\
(\g_1^{ccg})_1 =\
 \frac{27}{4}
{},
\label{Gccg2l}
\\
(\g_1^{ccg})_2  = 
 \frac{3375}{32}
{+} 
-\frac{135}{16}\, n_f 
{},
\label{Gccg3l}
\\
(\g_1^{ccg})_3  =   
\frac{46945}{24} 
+\frac{6561}{8}  \sbz \zeta_{3}
+\frac{243}{8}  \sbz \zeta_{4}
-\frac{13095}{16}  \sbz \zeta_{5}
\\
\hspace{19mm}{+} \, n_f 
\Bigg(
-\frac{14675}{72} 
-\frac{177}{2}  \sbz \zeta_{3}
-\frac{99}{4}  \sbz \zeta_{4}
\Bigg)
\nonumber
\\
\nnb
\hspace{38mm}
{+} \, n_f^2
\Bigg(
-\frac{251}{54} 
+6  \sbz \zeta_{3}
\Bigg)
{},
\label{Gccg4l}
\end{gather}
\begin{gather}
\nnb
(\g_1^{ccg})_4  =   
\frac{112928171}{2048} 
+\frac{11577699}{256}  \sbz \zeta_{3}
+\frac{815103}{128}  \,\zeta_3^2
\\
\hspace{22mm}
-\frac{1539243}{256}  \sbz \zeta_{4}
-\frac{23404221}{256}  \sbz \zeta_{5}
\\
\nnb
\hspace{23mm}
+\frac{2241675}{256}  \,\zeta_{6}
+\frac{22895649}{1024}  \,\zeta_{7}
\nonumber
\\
\nnb
{+}  n_f\, 
\Bigg(
-\frac{10723195}{1024} 
-\frac{1042157}{128}  \sbz \zeta_{3}
-\frac{14361}{64}  \,\zeta_3^2
\\
\nnb
\hspace{27mm}
+\,\frac{62571}{128}  \sbz \zeta_{4}
+\frac{1137861}{128}  \sbz \zeta_{5}
\\
\nnb
\hspace{27mm}
+\,\frac{77775}{128}  \,\zeta_{6}
-\frac{59535}{64}  \,\zeta_{7}
\Bigg)
\\
\nnb
{+} \, n_f^2\,
\Bigg(
\frac{572723}{2304} 
+\frac{8105}{16}  \sbz \zeta_{3}
-\frac{3789}{32}  \sbz \zeta_{4}
-\frac{2109}{8}  \sbz \zeta_{5}
\Bigg)
\nonumber
\\
\nnb
\hspace{25.5mm}
{+} \, n_f^3\,
\Bigg(
-\frac{2989}{864} 
-\frac{5}{3}  \sbz \zeta_{3}
+6  \sbz \zeta_{4}
\Bigg)
{}.
\end{gather}
Note that the leading renormalon contribution $\approx n_f^i \, \as^{i+1}$  vanishes
{\em (in any gauge!)}  due to the Taylor theorem which states, in particular,  that   
$\gamma^{ccg}_1 \equiv 0$ in the  Landau gauge.

\subsection{Five-loop running of the quark field} 

\begin{gather}
\g_2 = -\sum_{i=0}^{\infty} \Big( \g_2 \Big)_i\, h^{i+1}
{},
\\
(\g_2)_0 = \frac{4}{3},
\\
(\g_2)_1 = \frac{94}{3}-\frac{4}{3}\, n_f,
\\
(\g_2)_2 =
\frac{24941}{36} 
-26  \sbz \zeta_{3}
-\frac{1253}{18}\, n_f
+ \frac{20}{27}\,n_f^2
{},
\\
(\g_2)_3 =
\frac{19684159}{1296} 
-\frac{67469}{162}  \sbz \zeta_{3}
\\
\nnb
\hspace{27mm}
+501  \sbz \zeta_{4}
-\frac{129380}{81}  \sbz \zeta_{5}
\\\nnb
{+} \, n_f 
\Bigg(
-\frac{53713}{24} 
-\frac{5306}{27}  \sbz \zeta_{3}
-54  \sbz \zeta_{4}
-\frac{160}{3}  \sbz \zeta_{5}
\Bigg)
\\\nnb
\hspace{22mm}
{+} n_f^2\,
\Bigg(
\frac{10483}{243} 
+\frac{208}{9}  \sbz \zeta_{3}
\Bigg)
+
 \frac{140}{243}\, n_f^3
{},
\end{gather}
\begin{gather}
(\g_2)_4 =
\frac{2798900231}{7776} 
+\frac{17969627}{864}  \sbz \zeta_{3}
\\
\nnb
+\frac{13214911}{648}  \,\zeta_3^2
+\frac{16730765}{864}  \sbz \zeta_{4}
-\frac{832567417}{3888}  \sbz \zeta_{5}
\\
\nnb
\hspace{21mm}
+\frac{40109575}{1296}  \,\zeta_{6}
+\frac{124597529}{1728}  \,\zeta_{7}
\nonumber
\\
\nnb
{+}  n_f \,
\Bigg(
-\frac{861347053}{11664} 
-\frac{274621439}{11664}  \sbz \zeta_{3}
\\
\nnb
\hspace{9mm}
+\frac{1960337}{972}  \,\zeta_3^2
+\frac{465395}{1296}  \sbz \zeta_{4}
+\frac{22169149}{5832}  \sbz \zeta_{5}
\\
\nnb
\hspace{25mm}
+\frac{1278475}{1944}  \,\zeta_{6}
+\frac{3443909}{216}  \,\zeta_{7}
\Bigg)
\nonumber\\
\nnb
{+} \, n_f^2\,
\Bigg(
\frac{37300355}{11664} 
+\frac{1349831}{486}  \sbz \zeta_{3}
-\frac{128}{9}  \,\zeta_3^2
\\
\hspace{10mm}
-\frac{27415}{54}  \sbz \zeta_{4}
-\frac{12079}{27}  \sbz \zeta_{5}
-\frac{800}{9}  \,\zeta_{6}
-\frac{1323}{2}  \,\zeta_{7}
\Bigg)
\nonumber\\\nnb
{+} \, n_f^3
\Bigg(
-\frac{114049}{8748} 
-\frac{1396}{81}  \sbz \zeta_{3}
+\frac{208}{9}  \sbz \zeta_{4}
\Bigg)
\nonumber\\\nnb
\hspace{45mm}
{+} \, n_f^4
\Bigg(
\frac{332}{729} 
-\frac{64}{81}  \sbz \zeta_{3}
\Bigg)
{}.
\end{gather}

\subsection{Five-loop running of the quark mass} 
\begin{gather}
\g_m = -\sum_{i=0}^{\infty} \Big( \g_m \Big)_i\, h^{i+1}
{},
\\
(\g_m)_0 =4, 
\\
(\g_m)_1 = 
\frac{202}{3}
{-}
\frac{20}{9} \,n_f 
\\
(\g_m)_3 = 1249
{-} \,n_f\, 
\Bigg(
\frac{2216}{27} 
+\frac{160}{3}  \,\zeta_3
\Bigg)
-\frac{140}{81}\, n_f^2\,
{},
\\
\lefteqn{(\g_m)_3= 
\frac{4603055}{162} 
+\frac{135680}{27}  \,\zeta_3
-8800  \,\zeta_5
}
\\
{+} \,n_f\, 
\Bigg(
-\frac{91723}{27} 
-\frac{34192}{9}  \,\zeta_3
+880  \,\zeta_4
+\frac{18400}{9}  \,\zeta_5
\Bigg)
\nonumber\\
\hspace{18mm}{+} \, n_f^2\,
\Bigg(
\frac{5242}{243} 
+\frac{800}{9}  \,\zeta_3
-\frac{160}{3}  \,\zeta_4
\Bigg)
\\
\nnb
\hspace{33mm}{+} \, n_f^3\,
\Bigg(
-\frac{332}{243} 
+\frac{64}{27}  \,\zeta_3
\Bigg)
\nnb
{},
\end{gather}

\begin{gather}
(\gamma_m)_{4} = 
\frac{99512327}{162} 
 + \frac{46402466}{243}  \sbz \zeta_{3}
\\ 
\hspace{27.6mm}
+ 96800  \,\zeta_3^2
 - \frac{698126}{9}  \sbz \zeta_{4}
\nonumber\\
\hspace{10mm}
 - \frac{231757160}{243}  \sbz \zeta_{5}
 + 242000  \,\zeta_{6}
 + 412720  \,\zeta_{7}
\nonumber\\
{+} \, n_f\, 
\nnb
\Bigg(
-\frac{150736283}{1458} 
 - \frac{12538016}{81}  \sbz \zeta_{3}
 - \frac{75680}{9}  \,\zeta_3^2
\\ 
\nnb
\hspace{27.6mm}
 + \frac{2038742}{27}  \sbz \zeta_{4}
 + \frac{49876180}{243}  \sbz \zeta_{5}
\\ 
\nnb
\hspace{27.6mm}
- \frac{638000}{9}  \,\zeta_{6}
 - \frac{1820000}{27}  \,\zeta_{7}
\Bigg)
\label{gm5}
\\
\nnb
{+} \, n_f^2\,
\Bigg(
\frac{1320742}{729} 
 + \frac{2010824}{243}  \sbz \zeta_{3}
 + \frac{46400}{27}  \,\zeta_3^2
\\
\nnb
\hspace{14mm}
 - \frac{166300}{27}  \sbz \zeta_{4}
 - \frac{264040}{81}  \sbz \zeta_{5}
 + \frac{92000}{27}  \,\zeta_{6}
\Bigg)
\nonumber
\\
{+} \,
\fbox{$
 n_f^3\,
\Bigg(
\frac{91865}{1458} 
 + \frac{12848}{81}  \sbz \zeta_{3}
 + \frac{448}{9}  \sbz \zeta_{4}
 - \frac{5120}{27}  \sbz \zeta_{5}
\Bigg)
$}
\nnb
\\
\nnb
\hspace{14mm}
+ \, \fbox{$
 n_f^4\,
\Bigg(
-\frac{260}{243} 
 - \frac{320}{243}  \sbz \zeta_{3}
 + \frac{64}{27}  \sbz \zeta_{4}
\Bigg)
{}.
$}
\nonumber
\end{gather}

Note that in four-loop order we exactly\footnote{This agreement can be also considered as an important check of all our setup which is completely different from the ones utilized at the four-loop calculations.}  reproduce well-known results obtained in 
\cite{Chetyrkin:1997dh,Vermaseren:1997fq}.
The boxed terms in \re{gm5} are in full agreement with the results
derived previously on the basis of the $1/n_f$ method in
\mbox{\cite{PalanquesMestre:1983zy,Ciuchini:1999wy,Ciuchini:1999cv}.}

In  numerical form $\g_m$ reads
\begin{gather}
\nonumber
\lefteqn{\g_m =  - a_s - a_s^2 \left(4.20833 - 0.138889 n_f\right)}
\\ \nonumber
\hspace{5mm}
-
a_s^3  \left(19.5156 - 2.28412 n_f - 0.0270062 n_f^2 \right)  
\\ \nonumber 
\hspace{5mm}
-
a_s^4 \left(98.9434 - 19.1075 n_f 
\nonumber
\hspace{5mm}
\right.
\\
{}
\left.
 \hspace{13mm}+\, 0.276163 n_f^2  + 0.00579322 n_f^3 \right)
\\
\hspace{5mm}
-
a_s^5 \left(
559.7069 - 143.6864\, n_f + 7.4824\, n_f^2  
\nonumber
\right.
\\
{}
\left.
 \hspace{13mm}
+\, 0.1083 n_f^3  - 0.000085359\, n_f^4
\right)
{}.
\nonumber
\label{N[gm5qcd]}
\end{gather}

\begin{table}
\begin{center}
  \begin{tabular}{| c | c | c|c|c| }
    \hline
    $n_f$               &  3   & 4  &  5  &     6   \\ \hline
    $(\g_m)_4^{\rm exact}$       &  198.9   &  111.6  &  41.8  &     -9.8 \\ [0.51mm]    \hline
    \rule{0mm}{5.5mm}
    $\frac{1}{4^5}(\g_m)_4^{\rm APAP}$  \cite{Ellis:1997sb}         &  162.0   & 67.1  &  -13.7  &   -80.0 \\[1mm] 
    $\frac{1}{4^5}(\g_m)_4^{\rm APAP}$  \cite{Elias:1998bi}         &  163.0   & 75.2  &  12.6  &   12.2 \\[1mm]  
   $\frac{1}{4^5}(\g_m)_4^{\rm APAP}$   \cite{Kataev:2008ym}         &  164.0   & 71.6 &  -4.8  &   -64.6 \\ [1mm] 
    \hline
  \end{tabular}
\end{center}
\caption{The exact results for $(\g_m)_4$ together with the predictions made with the help of
the original APAP method and   its two somewhat  modified versions.}
\end{table}

Let us compare our numerical result for $(\g_m)_4$
\begin{gather}
\label{gm4_num}
\frac{1}{4^5}(\g_m)_4  =  559.71 - 143.6\, n_f 
+ 7.4824\, n_f^2
\\
\nnb
\hspace{20mm} + 0.1083\, n_f^3  - 0.00008535\, n_f^4 
{}, 
\end{gather}
with an  old prediction based on the 
``Asymptotic P\'ade Approximants'' (APAP) method \cite{Ellis:1997sb} (the boxed term  below 
was used as the  input)
\bga 
\frac{1}{4^5}(\g_m)_4^{\rm APAP}   =  530 - 143\, n_f + 6.67\, n_f^2 
\\
\nnb
\hspace{20mm} + 0.037\, n_f^3   -\fbox{$ 0.00008535\,n_f^4$} 
\end{gather}

However, this  good agreement is  broken  
for fixed values of $n_f$ due to severe cancellations between different powers of $n_f$ as one can see from the Table 1.

\section{Phenomenological applications of  $\gamma_m$ \label{gamma:m:pheno}}

\subsection{RG invariant quark mass}

The solution of eq. \re{anom-mass-def} reads:
\beq
\frac{m(\mu)}{m(\mu_0)} = \frac{c(a_s(\mu))}{c(a_s(\mu_0))}, \  \ \ 
c(x) = \mathrm{exp}\Biggl\{ \int {d x'} \frac{\g_m(x'}{\beta(x')} \Biggr\} 
{},
\label{cfun:1}
\eeq
\bga
c(x) =  (x)^{\bar{\g_0}} \Bigg\{ 1 + d_1 x + (d_1^2/2 + d_2) \,x^2 
\\          +  (d_1^3/6 + d_1 d_2 + d_3)\,  {x^3} \nonumber 
 +         (d_1^4/24 + d_1^2 d_2/2 
\\
+ d_2^2/2+ d_1 d_3 + d_4)
\,{x^4} + {\cal O}(x^5) \Bigg\}
\label{cfun:2}
{},
\end{gather}

\begin{flalign}
& d_1 = -\bar{\beta}_1 \, \bar{\gamma}_0 + \bar{\gamma}_1 {},&
\\
&d_2 = \bar{\beta}_1^2 \, \bar{\gamma}_0/2 - \bar{\beta}_2 \, \bar{\gamma}_0/2 - \bar{\beta}_1 \, \bar{\gamma}_1/2 + \bar{\gamma}
_2/2
{}, &
\\
&d_3 = -\bar{\beta}_1^3 \, \bar{\gamma}_0/3 + 2 \, \bar{\beta}_1 \, \bar{\beta}_2 \, \bar{\gamma}_0/3 - \bar{\beta}_3 \, 
\bar{\gamma}_0/3
\\
& \hspace{11mm}
 + \bar{\beta}_1^2 \, \bar{\gamma}_1/3 
          -\bar{\beta}_2 \, \bar{\gamma}_1/3 - \bar{\beta}_1 \, \bar{\gamma}_2/3 + \bar{\gamma}_3/3
{},
\nnb
\\
&d_4 = \bar{\beta}_1^4 \, \bar{\gamma}_0/4 - 3 \, \bar{\beta}_1^2 \, \bar{\beta}_2 \, \bar{\gamma}_0/4 + \bar{\beta}_2^2 \, \bar{
\gamma}_0/4 
\label{d4}
\\
& \hspace{18mm} + \bar{\beta}_1 \, \bar{\beta}_3 \, \bar{\gamma}_0/2 
    -  \bar{\beta}_4 \, \bar{\gamma}_0/4 - \bar{\beta}_1^3 \, \bar{\gamma}_1/4 
\nonumber
   \\ 
   & \hspace{18mm} + \bar{\beta}_1 \,  \bar{\beta}_2 \, \bar{\gamma}_1/2 - \bar{\beta}_3 \, \bar{\gamma}_1/4 +   \bar{\beta}_1^2 \
, \bar{\gamma}_2/4 
\nnb
\\
& \hspace{18mm}
- \bar{\beta}_2 \, \bar{\gamma}_2/4 
    - \bar{\beta}_1 \, \bar{\gamma}_3/4 + \bar{\gamma}_4/4 {}.
\nnb
&
\end{flalign}

Here 
$\bar{\g_i} = (\g_m)_i/\beta_0$, $\bar{\beta}_i = \beta_i/\beta_0$   
 and 
\[\beta(a_s)  =-\sum_{i \ge 0} \,\beta_i \, a_s^{i+2} = -\beta_0 \left\{\sum_{i \ge 0} \,\bar{\beta_i}\,  a_s^{i+2}
\right\}
\]
is the QCD $\beta$-function. Unfortunately, the coefficient $d_4$ in
eq.~\re{d4} does depend on the yet unknown {\em five-loop} coefficient
$\beta_4$ (up to four loops the $\beta$-function is known from
\cite{Gross:1973id,Politzer:1973fx,Caswell:1974gg,Jones:1974mm,Egorian:1978zx,Tarasov:1980au,Larin:1993tp,vanRitbergen:1997va,Czakon:2004bu}).

Numerically, the $c$-function  reads:
\bga
c(x)\bbuildrel{=\!=\!=}_{n_f = 3}^{}  x^{4/9}\,c_s(x), \ \
c(x)\bbuildrel{=\!=\!=}_{n_f = 4}^{}  x^{12/25}\,c_c(x),\ 
\nnb
\\
c(x)\bbuildrel{=\!=\!=}_{n_f = 5}^{}  x^{12/23}\,c_b(x), \ \
c(x)\bbuildrel{=\!=\!=}_{n_f = 6}^{}  x^{4/7}\,c_t(x)
{},
\nnb
\end{gather}
with
\bga
c_s(x) = 1  + 0.8950 \,x + 1.3714 \,x^2  \\ 
\hspace{7mm}+ 1.9517 \,x^3 + (15.6982-  0.11111\,\bar{\beta_4}) \,x^4    ,   
\nnb
\\
c_c(x) = 1 + 1.0141 \,x + 1.3892 \,x^2  
\\
\hspace{7mm}
+ 1.0905 \,x^3  + ( 9.1104  -  0.12000  \,\bar{\beta_4}) \,x^4     ,
\nonumber
\\
c_b(x) = 1+ 1.1755 \,x + 1.5007 \,x^2  
\\
\hspace{7mm}
+ 0.17248 \,x^3  + (  2.69277 -  0.13046  \,\bar{\beta_4}) \,x^4 ,
\nonumber
\\
c_t(x) = 1+  1.3980 \,x + 1.7935 \,x^2 
\\
\nnb
\hspace{7mm}
 - 0.68343 \,x^3  + ( - 3.5130 -    0.14286 \,\bar{\beta_4}) \,x^4   
\label{cfunctions}
{}.
\end{gather}
Eq. \re{cfun:1} could be used to define  the  important concept of  the  RGI mass
\beq
 {m}^{\rm RGI} \equiv m(\mu_0)/{c(a_s(\mu_0))}
\label{RGI}
{}, 
\eeq
A remarkable   property of the RGI mass is $\mu$ and 
{\em scheme} independency: in  {\em any} (mass-independent) scheme
\[ \lim_{\mu \to \infty} a_s(\mu)^{-\bar{\g}_0}\, \, m(\mu) = {m}^{\rm RGI}
{}.
\]
Due to this property  the   RGI mass  and function $c(x)$ 
are  often used in  the context  of lattice simulations. 
For example,  the   lattice {\bf ALPHA}collaboration uses \re{RGI}
to find the $\ovl{\mbox{MS}}$ mass of the strange quark
at a lower scale, say $m_s(2 \, \mbox{GeV})$,  from the $m_s^{\rm RGI}$ mass
determined from lattice simulations (see, e.g. \cite{DellaMorte:2005kg}). 
For example, setting $a_s(\mu = 2\, \mbox{GeV}) = \frac{\large\alpha_s(\mu)}{ \pi}
= 0.1$, we arrive at (here the formal parameter $h=1$ counts loops):
\begin{multline}
 m_s(2 \,\mbox{GeV}) =  m_s^{\rm RGI} \left(a_s(2\, \mbox{GeV}) \right)^{\frac{4}{9}} \times
\\
\Bigl(
1  + 0.0895\, h^2 + 0.0137\, h^3  + 0.00195\, h^4 
\\
 + (0.00157   - 0.000011 \,\ovl{\be}_4)\, h^5
\Bigr)
{}.
\label{cs2GeV}
\end{multline}
In order to have an idea of  effects due to the  five-loop term 
in \re{cs2GeV} one should make a guess about $\bar{\beta_4}$. By inspecting
the available four-loop result
\begin{gather}
\beta(n_f=3) = -\left(\frac{4}{9}\right)  \times
\\
\nnb
\Bigl(
\as + 1.777\, \as^2 + 4.4711\, \as^3 + 20.990\, \as^4 + \bar{\beta}_4\, \as^5
\Bigr)
{},
\end{gather}
we  conclude that $\ovl{\beta}_4 = 50-100$ could serve as a natural estimate of  $\ovl{\beta}_4$.
With this  choice we conclude  that the (apparent) convergence of
the above series is quite good even at a rather small energy scale of  2 GeV.
On the other hand, the authors of \cite{Elias:1998bi} cite an estimation 
$\bar{\beta}_4= -850(!)$ for  the $n_f=3$
QCD. With such  a huge   value of $\bar{\beta}_4$ the 
five loop term in \re{cs2GeV} would amount to $0.01092$ and, thus,  would significantly exceed
the four-loop contribution (0.00195).

\subsection{Higgs decay into quarks}

The (inclusive)  decay width of the Higgs boson into  a
pair of quarks is related to the spectral density of the scalar correlator
according to   (\ref{H2ff}-\ref{Pi:S}).
For   $n_f =5$ which  corresponds to the newly discovered Higgs boson we get
\bga
R^{S}(s = M_H^2,\mu=M_H) = 1 + 5.667\,  a_s
\\
\hspace{20mm}+ 29.147 \, a_s^2  
+  41.758 \,a_s^3 \,  {- 825.7}\,a_s^4
\nonumber
\\
=1 + 0.2041  + 0.0379  + 0.0020  {-0.00140}
\nonumber
\label{RS_as4_nl5}
{},
\end{gather}
where we set
$a_s = \alpha_s/\pi= 0.0360$ (for the Higgs mass value $M_H = 125$  GeV and $\alpha_s(M_Z) = 0.118$).
The decay rate   \re{H2ff} depends on two phenomenological parameters
 $\alpha_s(M_H)$ and the quark running mass $m_q$.  Let us 
consider, for definiteness, the dominant decay mode $H
\to {\bar b}{b}$.  To avoid the appearance of large logarithms of the type
$\ln \mu^2/M_H^2$ the parameter $\mu$ should be chosen 
around $M_H$.  However, the starting value of $m_b$ is usually
determined at a much smaller scale (typically around 5-10 GeV
\cite{Chetyrkin:2009fv}). The evolution of $m_b(\mu)$ from a lower
scale to $\mu = M_h$ is governed by eqs. (\ref{cfun:1}-\ref{d4})
which depend on 
the quark mass anomalous dimension
$\gamma(\als)$ and the QCD beta function $\beta(\als)$ (for QCD with
$n_f=5$).  In order to match the ${\cal O}(\als^4)$ accuracy of
\re{RS_as4_nl5} one should know {\em both} RG functions $\beta$ and
$\gamma_m $ in the five-loop approximation.

Let us  assume, conservatively, that $ 0 \le \, {\bar{\beta}_4}^{n_f=5} \le 200$.
The value of $m_b(\mu=M_H)$ is  obtained with RG running from
$m_b(\mu=\,10 \, \mbox{GeV})$ and, thus, depends on $\beta$ and
$\g_m$.  Using the Mathematica package RunDec\footnote{We have
extended the package by including the five-loop effects to the running
of $\als$ and quark masses.}  \cite{Chetyrkin:2000yt} and eq. 
\re{cfunctions} we  find for  the shift from  the five-loop term
\bga
\frac{\delta m_b^2(M_H)}{m_b^2(M_H)} =
 -1.3\cdot 10^{-4}({\bar{\beta}_4 =0}) 
\\
|-4.3 \cdot 10^{-4}({\bar{\beta}_4 =100})
| -7.3\cdot 10^{-4}({\bar{\beta}_4 =200})
\end{gather}
If we set $\mu=M_H$,  then the total  effect of ${\cal O}(\alpha_s^4$)
terms as coming from the five-loop
running and four-loop contribution to $R^S$ on 
$\Gamma(H \to \bar{b} b )$
would  be  around {-2\permil} (for $\bar{\beta}_4 =100$).  This is to be 
compared  to the parametric uncertainties  coming from the input
parameters $\alpha_s(M_Z) = 0.1185(6)$ \cite{Beringer:1900zz} and
$m_b(m_b)= 4.169(8)\ \mbox{GeV}$ \cite{Penin:2014zaa} which correspond
to {$\pm$ 1\permil}\ and {$\pm$ 4\permil}\ respectively.


\section{Deep Inenalsic Scattering  (DIS)}

\subsection{ Bjorken sum rule}
The Bjorken sum rule  (for polarized DIS)
expresses
the integral over the  spin distributions of quarks inside of the nucleon in terms of 
its axial charge times a  coefficient function
${C}^{Bjp}$:
\bea
\Gamma_1^{p-n}(Q^2) &=& 
\int_0^1 [g_1^{ep}(x,Q^2)-g_1^{en}(x,Q^2)]dx
\nnb
\\
&=&\frac{g_A}{6}
C^{Bjp}(a_s) +
\sum_{i=2}^{\infty}\frac{\mu_{2i}(Q^2)}{Q^{2i-2}}
{},
\label{gBSR}
\eea
where $g_1^{ep}$ and $g_1^{en}$ are the spin-dependent proton and neutron
structure functions, $g_A$ is the nucleon axial charge as measured in 
neutron $\beta$-decay. 
The coefficient function $C^{Bjp}(a_s) = 1 +{\cal O}(a_s)$ is  fixed by the  OPE
of two EM currents   (for a more detailed  discussion, see  \cite{Larin:2013yba}):
\begin{gather}
\int T{[ J_{\alpha}^{E}(x)J_{\beta}^{E}(0)]}e^{iqx}dx|_{q^2\rightarrow{-\infty}}
\approx  
\label{OPE}
\\
\hspace{25mm}
\frac{q^{\sigma}}{q^2}  \epsilon_{\alpha\beta\rho\sigma}
C^{Bjp}_{a}(a_s)\, A_{\rho}^{c}(0)+\dots
{},
\nnb
\end{gather}
\begin{gather}
C^{Bjp}_{a}(a_s) = 
\label{C:def}
\\
\nnb
\mathrm{Tr}[E^2 t_a]\, C^{Bjp}_{NS}(a_s) + \mathrm{Tr}(E)\,  \mathrm{Tr}[E t_a]\,C^{Bjp}_{SI}(a_s) =
\\
\nnb
\Bigg(  
C^{Bjp}_{NS}(a_s)  + 3\, \mathrm{Tr}[E]\  C^{Bjp}_{SI}(a_s)
\Bigg)
\,\, \mathrm{Tr}[E^2 t_a]
\nnb
\\
\nnb
\hspace{35mm}
 =C^{Bjp}(a_s)\, \mathrm{Tr}[E^2 t_a]
{}.
\end{gather}
Here  $E=diag(Q_i)$ is the quark  charge  matrix, 
$J_{\alpha}^{E} =  \ovl{\psi} E \gamma_{\rho} \psi$ is the quark EM current, 
$ A_{\rho}^{c} = \ovl{\psi} \, \gamma_{\rho} \,t^c\gamma_5  \,\psi$ is  
the (flavor non-singlet!) axial current and $Q^2=-q^2$.

\begin{center}
\begin{figure}
\includegraphics[width=23mm, angle=-90]{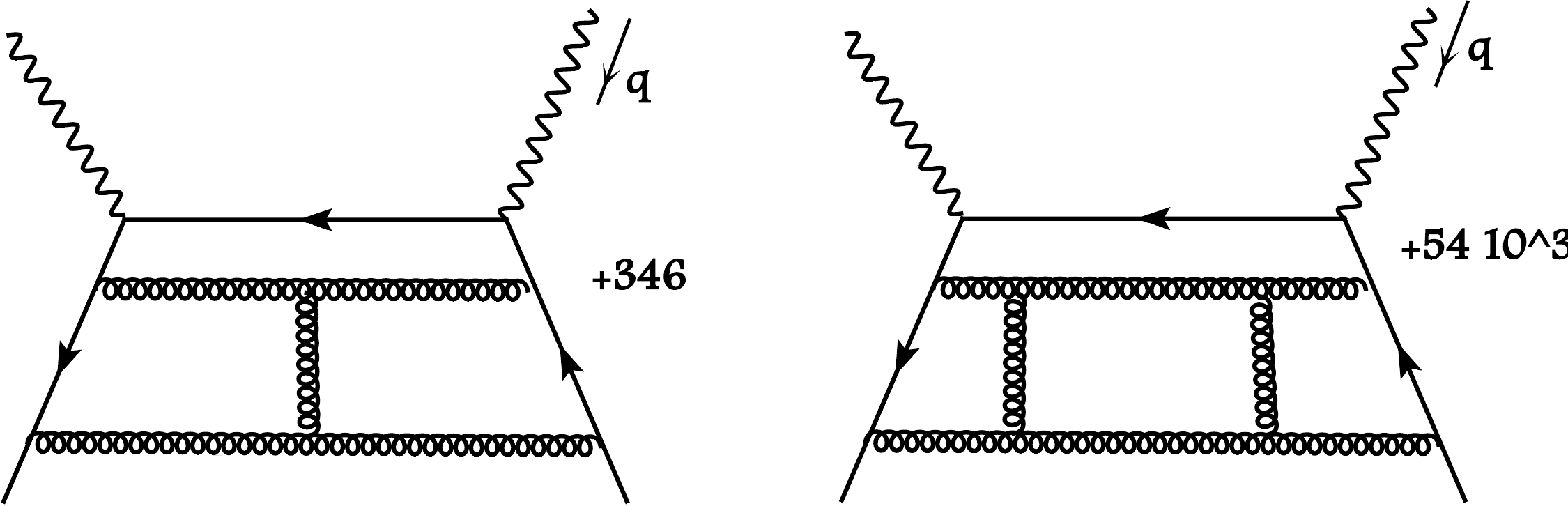}
\caption{Examples of diagrams contributing to  the coefficient function  $C^{Bjp}_{NS}$   at three
and four loops. }
\label{CBjp:NS}
\end{figure}
\end{center}

\begin{center}
\begin{figure}
\includegraphics[width=23mm, angle=-90]{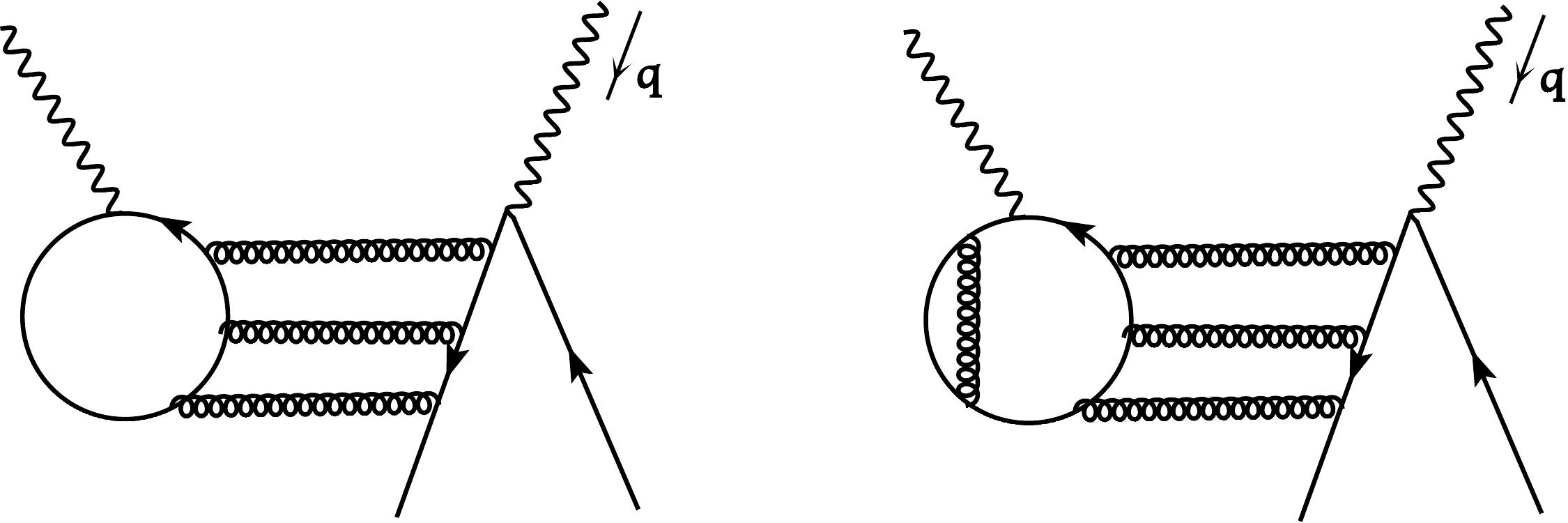}
\caption{Examples of diagrams contributing to  the coefficient function  $C^{Bjp}_{SI}$   at three
and four loops. }
\label{CBjp:SI}
\end{figure}
\end{center}

As one can see from eqs.~\re{C:def},   the coefficient function $C^{Bjp}(a_s)$  receives contributions from
two types of diagrams, viz.  the singlet  and  non-singlet ones (see Figs. \ref{CBjp:SI} and  \ref{CBjp:NS}
respectively). Let us discuss them in turn.

The coeffcient function $C^{Bjp}_{NS}(a_s)$ starts from 1 which
corresponds to the parton approximation. The four-loop result was
published in \cite{Baikov:2010je} for the case of a general gauge
group. The QCD result reads:

\begin{align}
& \hspace{-6mm} C^{Bjp}_{NS}(a_s) =~ 1 - a_s + a_s^2 \, \Big(-\frac{55}{12} + \frac{n_f}{3}\Big)
\label{CBjpas4}
\\[1mm]
\nn
& \hs~+ a_s^3\,\Bigg[-\frac{13841}{216} 
-\frac{44}{9}  \sbz \zeta_{3}
+\frac{55}{2}  \sbz \zeta_{5}
\\
\nonumber 
& \hs\qquad  + n_f\,\Big( \frac{10339}{1296} 
+\frac{61}{54}  \sbz \zeta_{3}
-\frac{5}{3}  \sbz \zeta_{5}\Big) - n_f^2\,\frac{115}{648}
\Bigg]
\nonumber
\nonumber \\[1mm]
& \hs~+ a_s^4\,\Bigg[
-\frac{17865665}{20736} 
+\frac{8213}{48}  \sbz \zeta_{3}
-\frac{363}{8}  \,\zeta_3^2
\nnb
\\
\nonumber 
&\hspace{1cm} \hs\qquad
+\frac{343175}{864}  \sbz \zeta_{5}
-\frac{2695}{16}  \,\zeta_{7}
\nonumber 
\\[1mm]
& \hs\qquad
 + n_f\,\Big(
\frac{10134475}{62208} 
-\frac{32743}{2592}  \sbz \zeta_{3}
+\frac{11}{2}  \,\zeta_3^2
\nnb
\\
&\hspace{1cm}
 \hs\qquad
-\frac{53215}{1296}  \sbz \zeta_{5}
+\frac{245}{24}  \,\zeta_{7}
        \Big)
\nonumber\\[1mm]
& \hs\qquad
 + n_f^2\,\Big(
-\frac{169523}{20736} 
+\frac{103}{432}  \sbz \zeta_{3}
-\frac{1}{6}  \,\zeta_3^2
+\frac{5}{12}  \sbz \zeta_{5}
\Big)
\nnb
\\
\nnb
&
\hspace{4cm} +
n_f^3\, \frac{605}{5832}
\Bigg]
{},
\end{align}

\begin{align}
& \hs {}C^{Bjp}_{NS} =
1 - a_s
{+} a_s^2\,
\Big(
-4.583
+0.3333  \,n_f
\Big)
\label{CBJN}
\\
& \hs \qquad {+}\, a_s^3\,
\Big(
-41.44
+7.607  \,n_f
-0.1775  \, n_f^2
\Big)
\nonumber\\
& \hs \qquad {+}\, \,a_s^4\,
\Big(
-479.4
+123.4  \,n_f
-7.697  \, n_f^2
+0.1037  \, n_f^3
\big)
\nonumber
{}.
\end{align}
Note that for phenomenologically
relevant values  of $Q^2 \le  3 \,  \mbox{GeV}^2$
one should  work in an effective QCD with  three active flavours. In this case the 
singlet contrubution vanishes identically as   $\mathrm{Tr}[E] \equiv 0$ and  
\begin{gather}
C^{Bjp}(n_f=3) \equiv  C^{Bjp}_{NS}(n_f=3)  =
\label{CBJNnf3}
\\
\hspace{13mm}
1
- a_s
-3.583  a_s^2
-20.22  a_s^3
 -175.7  \,a_s^4
{}.
\
\nnb
\end{gather}

\begin{center}
\begin{figure}
\hspace{3mm}\includegraphics[width=76mm]{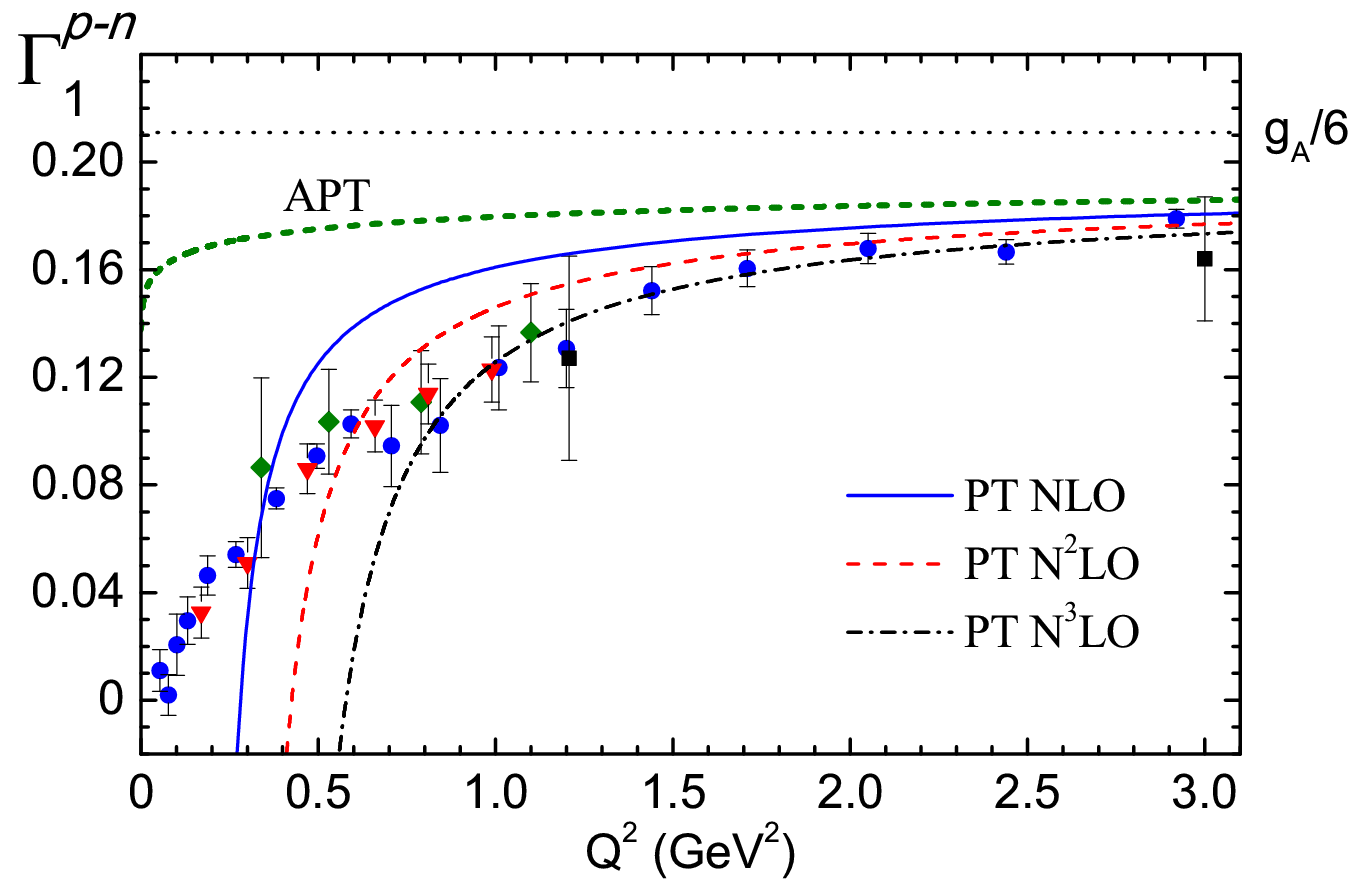}
\caption{\label{fig:PT_Gam}
Perturbative part of the Bjorken sum rule \re{gBSR}  as a function of the momentum
 transfer squared $Q^2$ in different orders  
 against the combined set of
 the Jefferson Lab (taken from 
V.L. Khandramai, R.S. Pasechnik, D.V. Shirkov, O.P. Solovtsova, O.V. Teryaev,
{\em Four-loop QCD analysis of the Bjorken sum rule vs data}, Phys.Lett.B706:340-344,2012).}
\label{Bjp:exp}
\end{figure}
\end{center}
Phenomelogical implications of \re{CBJNnf3} have been studied in works
\cite{Khandramai:2011zd,Khandramai:2013haz}. Their results  can be  summarized  as follows.

First, by comparing experimental 
data (see Fig.~\ref{Bjp:exp}) with the theoretical prediction
\re{CBJNnf3} the authors have arrived at  the following 
conclusion \cite{Khandramai:2011zd}:  "One
can see that at $Q^2 \ge 0.7 \ \mbox{GeV}^2$ the four-loop approximation
describes the data quite well. Moreover, the corresponding curve
passes close to the central values of several data points, although
the experimental accuracy (which is of the same order as both the
three- and four-loop contributions) does not allow one to make a
definite choice between four- and three-loop approximations."

Second, a certain {\em duality} between higher orders and higher-twist
contributions to the Bjorken sum rule \re{gBSR} has been
detected\footnote{The phenomenon in a more general context was earlier
discussed in \cite{Narison:2009ag}.} in
\cite{Khandramai:2013haz}. Indeed, the fitted value of first non-zero
higher twist contribution, $\mu_4$, has proved to be strongly
dependent on the order of PT terms kept in \re{CBJNnf3}. For example,
at leading order (that is with all terms in \re{CBJNnf3} except first
two set to zero) $\mu_4$ was found to be $-0.037 \pm 0.003 \
\mbox{GeV}^2$.  At next-leading-order $\mu_4$ is decreased to $-0.025
\pm 0.04 \ \mbox{GeV}^2$ and, finally, at ${\cal O}(\alpha_s^4)$ it
becomes compatible to zero: $\mu_4 = 0.005 \pm 0.008 \ \mbox{GeV}^2$.

The singlet contributions to $C^{Bjp}$ formally starts at two loops but
the corresponding diagrams sums to identical zero due to Furry's
Theorem. The next, three-loop term also happen to be zero \cite{Larin:1991tj}.  This fact
has been explained in \cite{Larin:2013yba} with   the help of generalized
Crewther relation \cite{Crewther:1972kn,Broadhurst:1993ru,Crewther:1997ux}. 
Here it has been also  predicted that at four loops
the singlet contribution should have the  form:
\beq
X \,  \beta_0 \  d^{abc}d^{abc}\,\, \Bigg (\frac{\alpha_s}{\pi} \Bigg)^4
{},
\label{larin}
\eeq
with $\beta_0=\frac{11}{12} C_A - \frac{T_f\,n_f}{3}$,
$d^{abc} = 2\, \mathrm{\mathrm{Tr}}(\{\frac{\lambda^a}{2},\frac{\lambda^b}{2}\}  \frac{\lambda^c}{2}\})$,
 and X  being a constant.

We have  performed a direct calculation of $C^{Bjp}_{SI}$ at  order $\alpha_s^4$.
Our result reads:
\[
C^{Bjp}_{SI} = \frac{1}{9} \, \beta_0 \  d^{abc}d^{abc}\, \Bigg (\frac{\alpha_s}{\pi} \Bigg)^4
\]
in full agreement\footnote{In fact, paper   \cite{Larin:2013yba} has also guessed a particular value 
of 
\mbox{$X= -\frac{1}{3}\Big(-\frac{179}{384} + \frac{25}{48}\zeta_3 - \frac{5}{24}\zeta_5$ \Big) }  
which happens to be  very different  from our result.}
with \re{larin}.


\section{Anomalous Dimensions of twist 2 operators}

Recently there has been a lot of progress in three-loop 
QCD calculations   of the moments (and the corresponding anomalous dimensions)
of deep inelastic structure functions  
\cite{Moch:2004pa,Vogt:2004mw,Moch:2004xu,Blumlein:2004xt,Vermaseren:2005qc}). 
In particular, 
the anomalous dimension of $\gamma_N^{NS}$ of the  twist-two non-singlet operator $\gamma^N_{NS}$
($\psi$ and $\psi'$ refers to the two {\em different}  quark species)
\beq
{\cal O}^{\{\mu_1,\dots,\mu_N\}} = 
\bar{\psi}'\,  \gamma^{\{\mu_1,} D^{\mu_2,\dots,\mu_N\}} \,\psi 
\label{Ot2n2}
\eeq
has been analytically  found for a   generic value of spin $N$
at the three loop level.  

In fact, a   general consistency argument requires the use of {\em four}-loop
splitting functions in applications of the results of 
\cite{Vermaseren:2005qc,Moch:2004xu} to the
phenomenological analysis of deep inelastic experimental data.
Unfortunately, the formidable problem of the {\em four-loop}
calculation for generic $N$ seems to be out of reach for  available
technologies. On the other hand,   fixed $N$ calculations are now possible (at least
for not too large values of $N$). 

The first result in this direction was reported in \cite{Baikov:2006ai} where
the four loop anomalous dimension of   the operator ${\cal O}^{\{\mu_1,\mu_2\}}$  was computed 
for a particular number of quark  species  $n_f=3$. Later this  result was confirmed and generalized
to a generic gauge  group in \cite{Velizhanin:2011es}. Note that at  four-loop level
the calculation of  $\gamma_N^{NS}$ with the use of BAICER  {\em does not require} 
application of any IRR:
one just  computes a diagonal  matrix element 
\[ \langle p| {\cal O}^{\{\mu_1,\dots,\mu_N\}}(0) | p \rangle, \]
with $|p \rangle$ being an off-shell  
quark state. {\em In principle} one could even compute the five-loop anomalous dimension  
$\gamma_N^{NS}$ for low   $N$. However, this would require significantly more computer as well as human  power 
(the  latter due to quite complicated IRR).

Two years ago  the present authors  computed $\gamma_N^{NS}$ for $N=2,3$ and $4$ in QCD
with full $n_f$ dependence\footnote{
The results given below were first presented on the 19 Meeting of SFB/TR9
``Computational Particle Physics'' 
19.03.2013 (Aachen). Very recently  $\gamma_3^{NS}$ and  $\gamma_4^{NS}$
have been   computed for  a case  of  a generic gauge  group \cite{Velizhanin:2014fua}.
For the QCD case  gauge we have found  full agreement between eqs. (\ref{t2n20}-\ref{t2n43})  and
results of   \cite{Velizhanin:2014fua}.
 }.   
Our results read:
\[\gamma_N^{NS} = \sum_{i \ge 0} \left(\gamma_N^{NS}\right)_i h^i {},\]
\begin{gather}
\left(\gamma_2^{NS}\right)_0 = \frac{32}{9}
\label{t2n20}
,
\\
\left(\gamma_2^{NS}\right)_1 =\frac{11744}{243} -  n_f\, \frac{256}{81}
,
\\
\left(\gamma_2^{NS}\right)_2 = 
- n_f^2\,\frac{896 }{729}+n_f \left(-\frac{1280}{27}\zt -\frac{167200}{2187}\right)
\nnb
\\
\hspace{27mm} +\frac{1280}{81}\zt +\frac{5514208}{6561}
,
\\
\left(\gamma_2^{NS}\right)_3 =  \frac{26060864}{6561}\zt -\frac{7040}{27}\zfr
\nnb
\\
\hspace{2cm}-\frac{1249280}{243}\zf+\frac{3100369144}{177147}
\nonumber\\
\hspace{11mm}+n_f\, \Biggl(-\frac{6322976}{2187}\zt+\frac{64640 }{81}\zfr
\\
\nnb
\hspace{2cm}+\frac{14720}{9}\zf-\frac{167219672}{59049}\Biggr)
\\
\nnb
\hspace{11mm}+n_f^2\, \Biggl(\frac{2560}{27}\zt -\frac{1280}{27}\zfr+\frac{1084904}{19683}\Bigg)
\\
\nnb
\hspace{11mm}+n_f^3 \,\Biggl(\frac{512}{243}\zt-\frac{4096}{6561}\Bigg)
{},
\end{gather}

\begin{gather}
\left(\gamma_3^{NS}\right)_0 =  \frac{50}{9}
,
\end{gather}
\begin{gather}
\left(\gamma_3^{NS}\right)_1 = \frac{17225}{243}
- \, n_f 
\,\frac{415}{81}
,
\end{gather}
\begin{gather}
\left(\gamma_3^{NS}\right)_2 = 
\frac{64486199}{52488} 
+\frac{1100}{81} 
\\
\hspace{11.0mm}
{-}  n_f\, 
\Biggl(
\frac{967495}{8748} 
+\frac{2000}{27}  \sbz \zeta_{3}
\Biggr)
- n_f^2\,\frac{2569}{1458}
\nnb
{},
\end{gather}
\begin{gather}
\left(\gamma_3^{NS}\right)_3 = 
\hspace{3.0mm} \frac{69231923065}{2834352} 
+\frac{73641835}{13122}  \sbz \zeta_{3}
\\
\nnb
\hspace{18.0mm}-\frac{6050}{27}  \sbz \zeta_{4}
-\frac{1834550}{243}  \sbz \zeta_{5}
\\
\nonumber
\hspace{11mm}+n_f\, \Biggl(
-\frac{1978909951}{472392} 
-\frac{9638360}{2187}  \sbz \zeta_{3}
\\
\nnb
\hspace{2cm}
+\frac{100100}{81}  \sbz \zeta_{4}
+\frac{23000}{9}  \sbz \zeta_{5}
  \Biggr)
\\
\nnb
\hspace{11mm}+n_f^2\, \Biggl(
\frac{1733306}{19683} 
+\frac{12200}{81}  \sbz \zeta_{3}
-\frac{2000}{27}  \sbz \zeta_{4}
\Bigg)
\\
\nnb
\hspace{11mm}+n_f^3 \,\Biggl(
-\frac{23587}{26244} 
+\frac{800}{243}  \sbz \zeta_{3}
\Bigg)
{},
\end{gather}
\begin{gather}
\left(\gamma_4^{NS}\right)_0 =  \frac{314}{45}
,
\end{gather}
\begin{gather}
\left(\gamma_4^{NS}\right)_1 =  \frac{2620957}{30375}
- \, n_f \frac{13271}{2025}
,
\end{gather}
\begin{gather}
\left(\gamma_4^{NS}\right)_2 = 
\frac{245787905651}{164025000} 
+\frac{5756}{405}  \sbz \zeta_{3}
\\
\nnb
\hspace{1mm} - n_f \,
\Biggl(\frac{726591271}{5467500} 
+\frac{2512}{27}  \sbz \zeta_{3}
\Biggr)
-n_f^2\,\frac{384277}{182250}
{},
\end{gather}
\begin{gather}
\left(\gamma_4^{NS}\right)_3 = 
\nnb
\frac{1267599127484293}{44286750000} 
+\frac{58681291019}{8201250}  \sbz \zeta_{3}
\\
\hspace{15mm}
-\frac{31658}{135}  \sbz \zeta_{4}
-\frac{32178794}{3645}  \sbz \zeta_{5}
\nonumber\\
\hspace{7mm}
+n_f\, \Biggl(
-\frac{7539856966909}{1476225000} 
-\frac{1495404568}{273375}  \sbz \zeta_{3}
\nnb
\\
\label{t2n43}
\hspace{18mm} 
+\frac{627476}{405}  \sbz \zeta_{4}
+\frac{1289656}{405}  \sbz \zeta_{5}
 \Biggr)
\\
\nnb
\hspace{7mm}+n_f^2\, \Biggl(
\frac{6771192712}{61509375} 
+\frac{8584}{45}  \sbz \zeta_{3}
-\frac{2512}{27}  \sbz \zeta_{4}
\Bigg)
\\
\nnb
\hspace{7mm}+n_f^3 \,\Biggl(
-\frac{17813699}{16402500} 
+\frac{5024}{1215}  \sbz \zeta_{3}
\Bigg)
{}.
\end{gather}

Numerically all 3 anomalous  dimensions  display a  remarkable similarity
(modulo global  normalization): 
\[\gamma^{NS}_{2} (n_f =3) =
\frac{32}{9\cdot 4}\left(\as + 2.7319 \as^2  + 7.8763 \as^3  + 28.706 \as^4\right)
{},
\]
\[\gamma^{NS}_{3} (n_f =3) =
\frac{50}{9\cdot 4}\left(\as + 2.4982 \as^2  + 7.0891 \as^3  + 23.587 \as^4\right)
{},
\]
\[\gamma^{NS}_{4} (n_f =3)=
\frac{314}{45\cdot 4}\left(\as + 2.3871 \as^2  + 6.8288 \as^3  + 22.294 \as^4\right)
{}.
\]
\section{Conclusions}

The problem of analytical evaluation of massless propagators at four
loops has been under investigation since long \cite{Baikov:2001aa}.  It has been
solved using reduction via $1/D$ expansion. As a result a number of
important four and five loop calculations have been done. In this
short review we have briefly discussed some of them related to the 
R-ratio, Higgs decays into quarks,   deep inelastic scattering and QCD renormalization group
functions.

\section{Acknowledgments}

This work was supported by
the Deutsche Forschungsgemeinschaft in the
Sonderforschungsbereich Transregio 9 ``Computational Particle Physics''. The work of P.~Baikov was 
supported in part by the Russian Ministry of Education and Science
under grant NSh-3042.2014.2.


\vspace{2cm}



\begin{thebibliography}{10}
\expandafter\ifx\csname url\endcsname\relax
  \def\url#1{\texttt{#1}}\fi
\expandafter\ifx\csname urlprefix\endcsname\relax\def\urlprefix{URL }\fi
\expandafter\ifx\csname href\endcsname\relax
  \def\href#1#2{#2} \def\path#1{#1}\fi

\bibitem{Smirnov:2002pj}
V.~A. Smirnov, Applied asymptotic expansions in momenta and masses, Springer,
  Berlin, 2002.

\bibitem{Stueckelberg53}
E.~Stueckelberg, A.~Petermann, {La normalisation des constantes dans la theorie
  des quanta}, Helv. Phys. Acta. 26 (1953) 499--520.

\bibitem{GellMann:1954fq}
M.~Gell-Mann, F.~Low, {Quantum electrodynamics at small distances}, Phys.Rev.
  95 (1954) 1300--1312.
\newblock \href {http://dx.doi.org/10.1103/PhysRev.95.1300}
  {\path{doi:10.1103/PhysRev.95.1300}}.

\bibitem{Bogolyubov:1956gh}
N.~Bogolyubov, D.~Shirkov, {Charge renormalization group in quantum field
  theory}, Nuovo Cim. 3 (1956) 845--863.
\newblock \href {http://dx.doi.org/10.1007/BF02823486}
  {\path{doi:10.1007/BF02823486}}.

\bibitem{Ashmore:1972uj}
J.~F. Ashmore, A method of gauge invariant regularization, Lett. Nuovo Cim. 4
  (1972) 289--290.

\bibitem{Cicuta:1972jf}
G.~M. Cicuta, E.~Montaldi, Analytic renormalization via continuous space
  dimension, Nuovo Cim. Lett. 4 (1972) 329--332.

\bibitem{tHooft:1972fi}
G.~'t~Hooft, M.~J.~G. Veltman, {Regularization and Renormalization of Gauge
  Fields}, Nucl. Phys. B44 (1972) 189--213.
\newblock \href {http://dx.doi.org/10.1016/0550-3213(72)90279-9}
  {\path{doi:10.1016/0550-3213(72)90279-9}}.

\bibitem{tHooft:1973mm}
G.~'t~Hooft, {Dimensional regularization and the renormalization group}, Nucl.
  Phys. B61 (1973) 455--468.
\newblock \href {http://dx.doi.org/10.1016/0550-3213(73)90376-3}
  {\path{doi:10.1016/0550-3213(73)90376-3}}.

\bibitem{Collins:1974da}
J.~C. Collins, {Normal Products in Dimensional Regularization}, Nucl. Phys. B92
  (1975) 477.
\newblock \href {http://dx.doi.org/10.1016/S0550-3213(75)80010-1}
  {\path{doi:10.1016/S0550-3213(75)80010-1}}.

\bibitem{Vladimirov:1979zm}
A.~A. Vladimirov, {Method For Computing Renormalization Group Functions In
  Dimensional Renormalization Scheme}, Theor. Math. Phys. 43 (1980) 417.
\newblock \href {http://dx.doi.org/10.1007/BF01018394}
  {\path{doi:10.1007/BF01018394}}.

\bibitem{Chetyrkin:1980pr}
K.~G. Chetyrkin, A.~L. Kataev, F.~V. Tkachov, {New Approach to Evaluation of
  Multiloop Feynman Integrals: The Gegenbauer Polynomial x Space Technique},
  Nucl. Phys. B174 (1980) 345--377.
\newblock \href {http://dx.doi.org/10.1016/0550-3213(80)90289-8}
  {\path{doi:10.1016/0550-3213(80)90289-8}}.

\bibitem{Chetyrkin:1984xa}
K.~G. Chetyrkin, V.~A. Smirnov, {$R^*$ Operation Corrected}, Phys. Lett. B144
  (1984) 419--424.
\newblock \href {http://dx.doi.org/10.1016/0370-2693(84)91291-7}
  {\path{doi:10.1016/0370-2693(84)91291-7}}.

\bibitem{ChetKuhn90}
K.~G. Chetyrkin, J.~H. K{\"u}hn, Mass corrections to the z decay rate, Phys.
  Lett. B248 (1990) 359--364.

\bibitem{Chetyrkin:1997qi}
K.~G. Chetyrkin, R.~Harlander, J.~H. K{\"u}hn, M.~Steinhauser, {Mass
  corrections to the vector current correlator}, Nucl. Phys. B503 (1997)
  339--353.
\newblock \href {http://arxiv.org/abs/hep-ph/9704222}
  {\path{arXiv:hep-ph/9704222}}, \href
  {http://dx.doi.org/10.1016/S0550-3213(97)00383-0}
  {\path{doi:10.1016/S0550-3213(97)00383-0}}.

\bibitem{Chetyrkin:2000zk}
K.~G. Chetyrkin, R.~V. Harlander, J.~H. K{\"u}hn, {Quartic mass corrections to
  $R_{had}$ at {${\cal O}(\alpha_s^3)$} }, Nucl. Phys. B586 (2000) 56--72.
\newblock \href {http://arxiv.org/abs/hep-ph/0005139}
  {\path{arXiv:hep-ph/0005139}}, \href
  {http://dx.doi.org/10.1016/S0550-3213(00)00393-X}
  {\path{doi:10.1016/S0550-3213(00)00393-X}}.

\bibitem{Baikov:2004ku}
P.~A. Baikov, K.~G. Chetyrkin, J.~H. K{\"u}hn, {Vacuum polarization in pQCD:
  First complete {${\cal O}(\alpha_s^4)$} result}, Nucl. Phys. Proc. Suppl. 135
  (2004) 243--246.
\newblock \href {http://dx.doi.org/10.1016/j.nuclphysbps.2004.09.013}
  {\path{doi:10.1016/j.nuclphysbps.2004.09.013}}.

\bibitem{Baikov:2009uw}
P.~Baikov, K.~Chetyrkin, J.~K{\"u}hn, {R(s) and hadronic tau-Decays in Order
  {$\alpha_s^4$}: Technical aspects}, Nucl.Phys.Proc.Suppl. 189 (2009) 49--53.
\newblock \href {http://arxiv.org/abs/0906.2987} {\path{arXiv:0906.2987}},
  \href {http://dx.doi.org/10.1016/j.nuclphysbps.2009.03.010}
  {\path{doi:10.1016/j.nuclphysbps.2009.03.010}}.

\bibitem{Blumlein:2012bf}
J.~Blumlein, {The Theory of Deeply Inelastic Scattering}, Prog.Part.Nucl.Phys.
  69 (2013) 28--84.
\newblock \href {http://arxiv.org/abs/1208.6087} {\path{arXiv:1208.6087}},
  \href {http://dx.doi.org/10.1016/j.ppnp.2012.09.006}
  {\path{doi:10.1016/j.ppnp.2012.09.006}}.

\bibitem{Gorishnii:1983su}
S.~G. Gorishny, S.~A. Larin, F.~V. Tkachov, {The Algorithm For OPE Coefficient
  Functions In The MS Scheme}, Phys. Lett. B124 (1983) 217--220.
\newblock \href {http://dx.doi.org/10.1016/0370-2693(83)91439-9}
  {\path{doi:10.1016/0370-2693(83)91439-9}}.

\bibitem{Gorishnii:1986gn}
S.~G. Gorishny, S.~A. Larin, {Coefficient Functions Of Asymptotic Operator
  Expansions In Minimal Subtraction Scheme}, Nucl. Phys. B283 (1987) 452.
\newblock \href {http://dx.doi.org/10.1016/0550-3213(87)90283-5}
  {\path{doi:10.1016/0550-3213(87)90283-5}}.

\bibitem{Larin:1991tj}
S.~A. Larin, J.~A.~M. Vermaseren, {The {$\alpha_s^3$} corrections to the
  Bjorken sum rule for polarized electroproduction and to the Gross-Llewellyn
  Smith sum rule}, Phys. Lett. B259 (1991) 345--352.

\bibitem{Smirnov:2006ry2}
V.~A. Smirnov, Feynman Integral Calculus, Springer, Berlin, 2006.

\bibitem{Smirnov:2012gma2}
V.~A. Smirnov, {Analytic tools for Feynman integrals}, Springer, Berlin, 2012.

\bibitem{Grozin:2007zz}
A.~Grozin, {Lectures on QED and QCD: Practical calculation and renormalization
  of one- and multi-loop Feynman diagrams}, Hackensack, USA: World Scientific, 2007.

\bibitem{Grozin:2011mt}
A.~Grozin, {Integration by parts: An Introduction}, Int.J.Mod.Phys. A26 (2011)
  2807--2854.
\newblock \href {http://arxiv.org/abs/1104.3993} {\path{arXiv:1104.3993}},
  \href {http://dx.doi.org/10.1142/S0217751X11053687}
  {\path{doi:10.1142/S0217751X11053687}}.



\bibitem{Smirnov:2010hn}
A.~Smirnov, A.~Petukhov, {The Number of Master Integrals is Finite},
  Lett.Math.Phys. 97 (2011) 37--44.
\newblock \href {http://arxiv.org/abs/1004.4199} {\path{arXiv:1004.4199}},
  \href {http://dx.doi.org/10.1007/s11005-010-0450-0}
  {\path{doi:10.1007/s11005-010-0450-0}}.

\bibitem{Lee:2013hzt}
R.~N. Lee, A.~A. Pomeransky, {Critical points and number of master integrals},
  JHEP 1311 (2013) 165.
\newblock \href {http://arxiv.org/abs/1308.6676} {\path{arXiv:1308.6676}},
  \href {http://dx.doi.org/10.1007/JHEP11(2013)165}
  {\path{doi:10.1007/JHEP11(2013)165}}.


\bibitem{Chetyrkin:1981qh}
K.~G. Chetyrkin, F.~V. Tkachov, {Integration by Parts: The Algorithm to
  Calculate beta Functions in 4 Loops}, Nucl. Phys. B192 (1981) 159--204.
\newblock \href {http://dx.doi.org/10.1016/0550-3213(81)90199-1}
  {\path{doi:10.1016/0550-3213(81)90199-1}}.



\bibitem{Laporta:1996rh}
S.~Laporta, E.~Remiddi, The analytic value of g(e)-2 at three loops in qed,
  Nucl. Phys. Proc. Suppl. 51C (1996) 142--147.

\bibitem{Laporta:2000dc}
S.~Laporta, Calculation of master integrals by difference equations, Phys.
  Lett. B504 (2001) 188--194.
\newblock \href {http://arxiv.org/abs/hep-ph/0102032}
  {\path{arXiv:hep-ph/0102032}}.

\bibitem{Baikov:2005nv}
P.~A. Baikov, A practical criterion of irreducibility of multi-loop feynman
  integrals, Phys. Lett. B634 (2006) 325--329.
\newblock \href {http://arxiv.org/abs/hep-ph/0507053}
  {\path{arXiv:hep-ph/0507053}}.

\bibitem{Baikov:1996rk}
P.~A. Baikov, {Explicit solutions of the 3--loop vacuum integral recurrence
  relations}, Phys. Lett. B385 (1996) 404--410.
\newblock \href {http://arxiv.org/abs/hep-ph/9603267}
  {\path{arXiv:hep-ph/9603267}}, \href
  {http://dx.doi.org/10.1016/0370-2693(96)00835-0}
  {\path{doi:10.1016/0370-2693(96)00835-0}}.

\bibitem{Grozin:2003ak}
A.~G. Grozin, {Lectures on multiloop calculations}, Int.J.Mod.Phys. A19 (2004)
  473--520.
\newblock \href {http://arxiv.org/abs/hep-ph/0307297}
  {\path{arXiv:hep-ph/0307297}}, \href
  {http://dx.doi.org/10.1142/S0217751X04016775}
  {\path{doi:10.1142/S0217751X04016775}}.

\bibitem{Baikov:2010hf}
P.~A. Baikov, K.~G. Chetyrkin, {Four-Loop Massless Propagators: an Algebraic
  Evaluation of All Master Integrals}, Nucl. Phys. B837 (2010) 186--220.
\newblock \href {http://arxiv.org/abs/1004.1153} {\path{arXiv:1004.1153}},
  \href {http://dx.doi.org/10.1016/j.nuclphysb.2010.05.004}
  {\path{doi:10.1016/j.nuclphysb.2010.05.004}}.

\bibitem{Smirnov:2010hd}
A.~V. Smirnov, M.~Tentyukov, {Four-Loop Massless Propagators: a Numerical
  Evaluation of All Master Integrals}, Nucl. Phys. B837 (2010) 40--49.
\newblock \href {http://arxiv.org/abs/1004.1149} {\path{arXiv:1004.1149}},
  \href {http://dx.doi.org/10.1016/j.nuclphysb.2010.04.020}
  {\path{doi:10.1016/j.nuclphysb.2010.04.020}}.

\bibitem{Lee:2011jt}
R.~N. Lee, A.~V. Smirnov, V.~A. Smirnov, {Master Integrals for Four-Loop
  Massless Propagators up to Transcendentality Weight Twelve}, Nucl. Phys. B856
  (2012) 95--110.
\newblock \href {http://arxiv.org/abs/1108.0732} {\path{arXiv:1108.0732}},
  \href {http://dx.doi.org/10.1016/j.nuclphysb.2011.11.005}
  {\path{doi:10.1016/j.nuclphysb.2011.11.005}}.

\bibitem{QGRAF}
P.~Nogueira, {Automatic Feynman graph generation}, J. Comput. Phys. 105 (1993)
  279--289.
\newblock \href {http://dx.doi.org/10.1006/jcph.1993.1074}
  {\path{doi:10.1006/jcph.1993.1074}}.

\bibitem{Vermaseren:2000nd}
J.~A.~M. Vermaseren, New features of form\href
  {http://arxiv.org/abs/math-ph/0010025} {\path{arXiv:math-ph/0010025}}.

\bibitem{Tentyukov:2004hz}
M.~Tentyukov, et~al., {ParFORM: Parallel Version of the Symbolic Manipulation
  Program FORM}\href {http://arxiv.org/abs/cs/0407066}
  {\path{arXiv:cs/0407066}}.

\bibitem{Tentyukov:2007mu}
M.~Tentyukov, J.~A.~M. Vermaseren, {The multithreaded version of FORM}\href
  {http://arxiv.org/abs/hep-ph/0702279} {\path{arXiv:hep-ph/0702279}}.

\bibitem{Baikov:2005rw}
P.~A. Baikov, K.~G. Chetyrkin, J.~H. K{\"u}hn, {Scalar correlator at {${\cal
  O}(\alpha_s^4)$}, Higgs decay into b- quarks and bounds on the light quark
  masses}, Phys. Rev. Lett. 96 (2006) 012003.
\newblock \href {http://arxiv.org/abs/hep-ph/0511063}
  {\path{arXiv:hep-ph/0511063}}, \href
  {http://dx.doi.org/10.1103/PhysRevLett.96.012003}
  {\path{doi:10.1103/PhysRevLett.96.012003}}.

\bibitem{Chetyrkin:2005kn}
K.~G. Chetyrkin, A.~Khodjamirian, {Strange Quark Mass from Pseudoscalar Sum
  Rule with ${\cal O}(\alpha_s^4)$ Accuracy}, Eur. Phys. J. C46 (2006)
  721--728.
\newblock \href {http://arxiv.org/abs/hep-ph/0512295}
  {\path{arXiv:hep-ph/0512295}}, \href
  {http://dx.doi.org/10.1140/epjc/s2006-02508-8}
  {\path{doi:10.1140/epjc/s2006-02508-8}}.

\bibitem{Gorishnii:1990zu}
S.~G. Gorishny, A.~L. Kataev, S.~A. Larin, L.~R. Surguladze, Corrected three
  loop qcd correction to the correlator of the quark scalar currents and gamma
  (tot) (h0 $\to$ hadrons), Mod. Phys. Lett. A5 (1990) 2703--2712.

\bibitem{Chetyrkin:1996sr}
K.~G. Chetyrkin, {Correlator of the quark scalar currents and $\Gamma_{\rm
  tot}(H \to \mbox{hadrons})$ at ${\cal O}(\alpha_s^3)$ in pQCD}, Phys. Lett.
  B390 (1997) 309--317.
\newblock \href {http://arxiv.org/abs/hep-ph/9608318}
  {\path{arXiv:hep-ph/9608318}}, \href
  {http://dx.doi.org/10.1016/S0370-2693(96)01368-8}
  {\path{doi:10.1016/S0370-2693(96)01368-8}}.

\bibitem{Chetyrkin:1979bj}
K.~G. Chetyrkin, A.~L. Kataev, F.~V. Tkachov, {Higher Order Corrections to
  {$\sigma_{tot}(e^+ e^- \to \mbox{Hadrons})$} in Quantum Chromodynamics},
  Phys. Lett. B85 (1979) 277.
\newblock \href {http://dx.doi.org/10.1016/0370-2693(79)90596-3}
  {\path{doi:10.1016/0370-2693(79)90596-3}}.

\bibitem{Gorishnii:1991vf}
S.~G. Gorishny, A.~L. Kataev, S.~A. Larin, The { ${\cal O}(\alpha_s^3)$}
  corrections to {$\sigma_{\rm tot}(e^+ e^- \to {\rm hadrons})$} and {$\sigma(
  {\tau} \to \nu_{\tau} + {\rm hadrons})$} in { QCD}, Phys. Lett. B259 (1991)
  144--150.

\bibitem{Baikov:2001aa}
P.~A. Baikov, K.~G. Chetyrkin, J.~H. K{\"u}hn, {The cross section of e+ e-
  annihilation into hadrons of order {$\alpha_s^4 n_f^2$} in perturbative QCD},
  Phys. Rev. Lett. 88 (2002) 012001.
\newblock \href {http://arxiv.org/abs/hep-ph/0108197}
  {\path{arXiv:hep-ph/0108197}}, \href
  {http://dx.doi.org/10.1103/PhysRevLett.88.012001}
  {\path{doi:10.1103/PhysRevLett.88.012001}}.

\bibitem{Baikov:2002uw}
P.~A. Baikov, K.~G. Chetyrkin, J.~H. K{\"u}hn, {Towards order {$\alpha_s^4$}
  accuracy in tau decays}, Phys. Rev. D67 (2003) 074026.
\newblock \href {http://arxiv.org/abs/hep-ph/0212299}
  {\path{arXiv:hep-ph/0212299}}, \href
  {http://dx.doi.org/10.1103/PhysRevD.67.074026}
  {\path{doi:10.1103/PhysRevD.67.074026}}.

\bibitem{Baikov:2002va}
P.~A. Baikov, K.~G. Chetyrkin, J.~H. K{\"u}hn, {Five-loop vacuum polarization
  in pQCD: {${\cal O}(m_q^2 \alpha_s^4 n_f^2)$} contribution}, Phys. Lett. B559
  (2003) 245--251.
\newblock \href {http://arxiv.org/abs/hep-ph/0212303}
  {\path{arXiv:hep-ph/0212303}}, \href
  {http://dx.doi.org/10.1016/S0370-2693(03)00186-2}
  {\path{doi:10.1016/S0370-2693(03)00186-2}}.

\bibitem{Baikov:2003gu}
P.~A. Baikov, K.~G. Chetyrkin, J.~H. K{\"u}hn, {QCD corrections to hadronic Z
  and tau decays}, Eur. Phys. J. C33 (2004) s650--s652.
\newblock \href {http://arxiv.org/abs/hep-ph/0311137}
  {\path{arXiv:hep-ph/0311137}}, \href
  {http://dx.doi.org/10.1140/epjcd/s2004-03-1839-8}
  {\path{doi:10.1140/epjcd/s2004-03-1839-8}}.

\bibitem{Baikov:2005sw}
P.~A. Baikov, K.~G. Chetyrkin, J.~H. K{\"u}hn, {Perturbative QCD and
  tau-decays}, Nucl. Phys. Proc. Suppl. 144 (2005) 81--87.
\newblock \href {http://dx.doi.org/10.1016/j.nuclphysbps.2005.02.011}
  {\path{doi:10.1016/j.nuclphysbps.2005.02.011}}.

\bibitem{Baikov:2006nb}
P.~A. Baikov, K.~G. Chetyrkin, J.~H. K{\"u}hn, {Multiloop calculations: Towards
  R at order $\alpha_s^4$}, Nucl. Phys. Proc. Suppl. 157 (2006) 27--31.
\newblock \href {http://arxiv.org/abs/hep-ph/0602126}
  {\path{arXiv:hep-ph/0602126}}, \href
  {http://dx.doi.org/10.1016/j.nuclphysbps.2006.03.005}
  {\path{doi:10.1016/j.nuclphysbps.2006.03.005}}.

\bibitem{Baikov:2008jh}
P.~A. Baikov, K.~G. Chetyrkin, J.~H. K{\"u}hn, {Order $\alpha^4_s$ QCD
  Corrections to $Z$ and $\tau$ Decays}, Phys. Rev. Lett. 101 (2008) 012002.
\newblock \href {http://arxiv.org/abs/0801.1821} {\path{arXiv:0801.1821}},
  \href {http://dx.doi.org/10.1103/PhysRevLett.101.012002}
  {\path{doi:10.1103/PhysRevLett.101.012002}}.

\bibitem{Baikov:2012zm}
P.~Baikov, K.~Chetyrkin, J.~K{\"u}hn, J.~Rittinger, {Vector Correlator in
  Massless QCD at Order ${\cal O}(\alpha_s^4)$ and the QED $\beta$-function at
  Five Loop}, JHEP 1207 (2012) 017.
\newblock \href {http://arxiv.org/abs/1206.1284} {\path{arXiv:1206.1284}},
  \href {http://dx.doi.org/10.1007/JHEP07(2012)017}
  {\path{doi:10.1007/JHEP07(2012)017}}.

\bibitem{Baikov:2010iw}
P.~Baikov, K.~Chetyrkin, J.~K{\"u}hn, {Adler Function, DIS sum rules and
  Crewther Relations}, Nucl.Phys.Proc.Suppl. 205-206 (2010) 237--241.
\newblock \href {http://arxiv.org/abs/1007.0478} {\path{arXiv:1007.0478}},
  \href {http://dx.doi.org/10.1016/j.nuclphysbps.2010.08.049}
  {\path{doi:10.1016/j.nuclphysbps.2010.08.049}}.

\bibitem{Baikov:2012er}
P.~Baikov, K.~Chetyrkin, J.~K{\"u}hn, J.~Rittinger, {Complete ${\cal
  O}(\alpha_s^4)$ QCD Corrections to Hadronic $Z$-Decays}, Phys.Rev.Lett. 108
  (2012) 222003.
\newblock \href {http://arxiv.org/abs/1201.5804} {\path{arXiv:1201.5804}},
  \href {http://dx.doi.org/10.1103/PhysRevLett.108.222003}
  {\path{doi:10.1103/PhysRevLett.108.222003}}.

\bibitem{Baikov:2012zn}
P.~Baikov, K.~Chetyrkin, J.~K{\"u}hn, J.~Rittinger, {Adler Function, Sum Rules
  and Crewther Relation of Order {${\cal O}(\alpha_s^4)$}: the Singlet Case},
  Phys.Lett. B714 (2012) 62--65.
\newblock \href {http://arxiv.org/abs/1206.1288} {\path{arXiv:1206.1288}},
  \href {http://dx.doi.org/10.1016/j.physletb.2012.06.052}
  {\path{doi:10.1016/j.physletb.2012.06.052}}.

\bibitem{Gorishnii:1991kd}
S.~G. Gorishny, A.~L. Kataev, S.~A. Larin, L.~R. Surguladze, The analytical
  four loop corrections to the qed beta function in the ms scheme and to the
  qed psi function: Total reevaluation, Phys. Lett. B256 (1991) 81--86.

\bibitem{Gross:1973id}
D.~J. Gross, F.~Wilczek, Ultraviolet behavior of non-abelian gauge theories,
  Phys. Rev. Lett. 30 (1973) 1343--1346.

\bibitem{Politzer:1973fx}
H.~D. Politzer, Reliable perturbative results for strong interactions?, Phys.
  Rev. Lett. 30 (1973) 1346--1349.

\bibitem{Caswell:1974gg}
W.~E. Caswell, Asymptotic behavior of nonabelian gauge theories to two loop
  order, Phys. Rev. Lett. 33 (1974) 244.

\bibitem{Jones:1974mm}
D.~R.~T. Jones, Two loop diagrams in yang-mills theory, Nucl. Phys. B75 (1974)
  531.

\bibitem{Egorian:1978zx}
E.~Egorian, O.~V. Tarasov, Two loop renormalization of the qcd in an arbitrary
  gauge, Theor. Math. Phys. 41 (1979) 863--867.

\bibitem{Tarasov:1980au}
O.~V. Tarasov, A.~A. Vladimirov, A.~Y. Zharkov, The gell-mann-low function of
  qcd in the three loop approximation, Phys. Lett. B93 (1980) 429--432.

\bibitem{Larin:1993tp}
S.~A. Larin, J.~A.~M. Vermaseren, The three loop qcd beta function and
  anomalous dimensions, Phys. Lett. B303 (1993) 334--336.
\newblock \href {http://arxiv.org/abs/hep-ph/9302208}
  {\path{arXiv:hep-ph/9302208}}.

\bibitem{vanRitbergen:1997va}
T.~van Ritbergen, J.~A.~M. Vermaseren, S.~A. Larin, The four-loop beta function
  in quantum chromodynamics, Phys. Lett. B400 (1997) 379--384.
\newblock \href {http://arxiv.org/abs/hep-ph/9701390}
  {\path{arXiv:hep-ph/9701390}}.

\bibitem{Czakon:2004bu}
M.~Czakon, {The four-loop QCD beta-function and anomalous dimensions}, Nucl.
  Phys. B710 (2005) 485--498.
\newblock \href {http://arxiv.org/abs/hep-ph/0411261}
  {\path{arXiv:hep-ph/0411261}}.

\bibitem{Tarrach:1980up}
R.~Tarrach, The pole mass in perturbative qcd, Nucl. Phys. B183 (1981) 384.

\bibitem{Tarasov:1982gk}
O.~V. Tarasov, Anomalous dimensions of quark masses in three loop
  approximationJINR-P2-82-900.

\bibitem{Larin:1993tq}
S.~A. Larin, The renormalization of the axial anomaly in dimensional
  regularization, Phys. Lett. B303 (1993) 113--118.
\newblock \href {http://arxiv.org/abs/hep-ph/9302240}
  {\path{arXiv:hep-ph/9302240}}.

\bibitem{Chetyrkin:1997dh}
K.~G. Chetyrkin, Quark mass anomalous dimension to {${\cal O}(\alpha_s^4)$},
  Phys. Lett. B404 (1997) 161--165.
\newblock \href {http://arxiv.org/abs/hep-ph/9703278}
  {\path{arXiv:hep-ph/9703278}}.

\bibitem{Vermaseren:1997fq}
J.~A.~M. Vermaseren, S.~A. Larin, T.~van Ritbergen, The 4-loop quark mass
  anomalous dimension and the invariant quark mass, Phys. Lett. B405 (1997)
  327--333.
\newblock \href {http://arxiv.org/abs/hep-ph/9703284}
  {\path{arXiv:hep-ph/9703284}}.

\bibitem{PalanquesMestre:1983zy}
A.~Palanques-Mestre, P.~Pascual, The 1/n-f expansion of the gamma and beta
  functions in qed, Commun. Math. Phys. 95 (1984) 277.

\bibitem{Ciuchini:1999wy}
M.~Ciuchini, S.~E. Derkachov, J.~Gracey, A.~Manashov, {Computation of quark
  mass anomalous dimension at O(1 / N**2(f)) in quantum chromodynamics},
  Nucl.Phys. B579 (2000) 56--100.
\newblock \href {http://arxiv.org/abs/hep-ph/9912221}
  {\path{arXiv:hep-ph/9912221}}, \href
  {http://dx.doi.org/10.1016/S0550-3213(00)00209-1}
  {\path{doi:10.1016/S0550-3213(00)00209-1}}.

\bibitem{Ciuchini:1999cv}
M.~Ciuchini, S.~E. Derkachov, J.~Gracey, A.~Manashov, {Quark mass anomalous
  dimension at O(1/N(f)**2) in QCD}, Phys.Lett. B458 (1999) 117--126.
\newblock \href {http://arxiv.org/abs/hep-ph/9903410}
  {\path{arXiv:hep-ph/9903410}}, \href
  {http://dx.doi.org/10.1016/S0370-2693(99)00573-0}
  {\path{doi:10.1016/S0370-2693(99)00573-0}}.

\bibitem{Ellis:1997sb}
J.~R. Ellis, I.~Jack, D.~Jones, M.~Karliner, M.~Samuel, {Asymptotic Pade
  approximant predictions: Up to five loops in QCD and SQCD}, Phys.Rev. D57
  (1998) 2665--2675.
\newblock \href {http://arxiv.org/abs/hep-ph/9710302}
  {\path{arXiv:hep-ph/9710302}}, \href
  {http://dx.doi.org/10.1103/PhysRevD.57.2665}
  {\path{doi:10.1103/PhysRevD.57.2665}}.

\bibitem{Elias:1998bi}
V.~Elias, T.~G. Steele, F.~Chishtie, R.~Migneron, K.~B. Sprague, {Pade
  improvement of QCD running coupling constants, running masses, Higgs decay
  rates, and scalar channel sum rules}, Phys.Rev. D58 (1998) 116007.
\newblock \href {http://arxiv.org/abs/hep-ph/9806324}
  {\path{arXiv:hep-ph/9806324}}, \href
  {http://dx.doi.org/10.1103/PhysRevD.58.116007}
  {\path{doi:10.1103/PhysRevD.58.116007}}.

\bibitem{Kataev:2008ym}
A.~Kataev, V.~Kim, {Higgs boson decay into bottom quarks and uncertainties of
  perturbative QCD predictions}\href {http://arxiv.org/abs/0804.3992}
  {\path{arXiv:0804.3992}}.

\bibitem{DellaMorte:2005kg}
M.~Della~Morte, et~al., Non-perturbative quark mass renormalization in
  two-flavor qcd, Nucl. Phys. B729 (2005) 117--134.
\newblock \href {http://arxiv.org/abs/hep-lat/0507035}
  {\path{arXiv:hep-lat/0507035}}.

\bibitem{Chetyrkin:2009fv}
K.~Chetyrkin, J.~K\"uhn, A.~Maier, P.~Maierhofer, P.~Marquard, et~al., {Charm
  and Bottom Quark Masses: An Update}, Phys.Rev. D80 (2009) 074010.
\newblock \href {http://arxiv.org/abs/0907.2110} {\path{arXiv:0907.2110}},
  \href {http://dx.doi.org/10.1103/PhysRevD.80.074010}
  {\path{doi:10.1103/PhysRevD.80.074010}}.

\bibitem{Chetyrkin:2000yt}
K.~G. Chetyrkin, J.~H. K{\"u}hn, M.~Steinhauser, Rundec: A mathematica package
  for running and decoupling of the strong coupling and quark masses, Comput.
  Phys. Commun. 133 (2000) 43--65.
\newblock \href {http://arxiv.org/abs/hep-ph/0004189}
  {\path{arXiv:hep-ph/0004189}}.

\bibitem{Beringer:1900zz}
J.~Beringer, et~al., {Review of Particle Physics (RPP)}, Phys.Rev. D86 (2012)
  010001.
\newblock \href {http://dx.doi.org/10.1103/PhysRevD.86.010001}
  {\path{doi:10.1103/PhysRevD.86.010001}}.

\bibitem{Penin:2014zaa}
A.~A. Penin, N.~Zerf, {Bottom Quark Mass from $\Upsilon$ Sum Rules to ${\cal
  O}(\alpha_s^3)$}\href {http://arxiv.org/abs/1401.7035}
  {\path{arXiv:1401.7035}}.

\bibitem{Larin:2013yba}
S.~Larin, {The singlet contribution to the Bjorken sum rule for polarized deep
  inelastic scattering}, Phys.Lett. B723 (2013) 348--350.
\newblock \href {http://arxiv.org/abs/1303.4021} {\path{arXiv:1303.4021}},
  \href {http://dx.doi.org/10.1016/j.physletb.2013.05.026}
  {\path{doi:10.1016/j.physletb.2013.05.026}}.

\bibitem{Baikov:2010je}
P.~A. Baikov, K.~G. Chetyrkin, J.~H. K{\"u}hn, {Adler Function, Bjorken Sum
  Rule, and the Crewther Relation to Order $\alpha_s^4$ in a General Gauge
  Theory}, Phys. Rev. Lett. 104 (2010) 132004.
\newblock \href {http://arxiv.org/abs/1001.3606} {\path{arXiv:1001.3606}},
  \href {http://dx.doi.org/10.1103/PhysRevLett.104.132004}
  {\path{doi:10.1103/PhysRevLett.104.132004}}.

\bibitem{Khandramai:2011zd}
V.~Khandramai, R.~Pasechnik, D.~Shirkov, O.~Solovtsova, O.~Teryaev, {Four-loop
  QCD analysis of the Bjorken sum rule vs data}, Phys.Lett. B706 (2012)
  340--344.
\newblock \href {http://arxiv.org/abs/1106.6352} {\path{arXiv:1106.6352}},
  \href {http://dx.doi.org/10.1016/j.physletb.2011.11.023}
  {\path{doi:10.1016/j.physletb.2011.11.023}}.

\bibitem{Khandramai:2013haz}
V.~Khandramai, O.~Solovtsova, O.~Teryaev, {Polarized Bjorken Sum Rule Analysis:
  Revised}, Nonlin.Phenom.Complex Syst. 16 (2013) 93--98.
\newblock \href {http://arxiv.org/abs/1302.3952} {\path{arXiv:1302.3952}}.

\bibitem{Narison:2009ag}
S.~Narison, V.~Zakharov, {Duality between QCD Perturbative Series and Power
  Corrections}, Phys.Lett. B679 (2009) 355--361.
\newblock \href {http://arxiv.org/abs/0906.4312} {\path{arXiv:0906.4312}},
  \href {http://dx.doi.org/10.1016/j.physletb.2009.07.060}
  {\path{doi:10.1016/j.physletb.2009.07.060}}.

\bibitem{Crewther:1972kn}
R.~J. Crewther, Nonperturbative evaluation of the anomalies in low-energy
  theorems, Phys. Rev. Lett. 28 (1972) 1421.

\bibitem{Broadhurst:1993ru}
D.~J. Broadhurst, A.~L. Kataev, Connections between deep inelastic and
  annihilation processes at next to next-to-leading order and beyond, Phys.
  Lett. B315 (1993) 179--187.
\newblock \href {http://arxiv.org/abs/hep-ph/9308274}
  {\path{arXiv:hep-ph/9308274}}.

\bibitem{Crewther:1997ux}
R.~J. Crewther, {Relating inclusive e+ e- annihilation to electroproduction sum
  rules in quantum chromodynamics}, Phys. Lett. B397 (1997) 137--142.
\newblock \href {http://arxiv.org/abs/hep-ph/9701321}
  {\path{arXiv:hep-ph/9701321}}, \href
  {http://dx.doi.org/10.1016/S0370-2693(97)00157-3}
  {\path{doi:10.1016/S0370-2693(97)00157-3}}.

\bibitem{Moch:2004pa}
S.~Moch, J.~A.~M. Vermaseren, A.~Vogt, The three-loop splitting functions in
  {QCD}: The non-singlet case, Nucl. Phys. B688 (2004) 101--134.
\newblock \href {http://arxiv.org/abs/hep-ph/0403192}
  {\path{arXiv:hep-ph/0403192}}.

\bibitem{Vogt:2004mw}
A.~Vogt, S.~Moch, J.~A.~M. Vermaseren, The three-loop splitting functions in
  {QCD}: The singlet case, Nucl. Phys. B691 (2004) 129--181.
\newblock \href {http://arxiv.org/abs/hep-ph/0404111}
  {\path{arXiv:hep-ph/0404111}}.

\bibitem{Moch:2004xu}
S.~Moch, J.~A.~M. Vermaseren, A.~Vogt, The longitudinal structure function at
  the third order, Phys. Lett. B606 (2005) 123--129.
\newblock \href {http://arxiv.org/abs/hep-ph/0411112}
  {\path{arXiv:hep-ph/0411112}}.

\bibitem{Blumlein:2004xt}
J.~Blumlein, J.~A.~M. Vermaseren, The 16th moment of the non-singlet structure
  functions f2(x,q**2) and f(l)(x,q**2) to o(alpha(s)**3), Phys. Lett. B606
  (2005) 130--138.
\newblock \href {http://arxiv.org/abs/hep-ph/0411111}
  {\path{arXiv:hep-ph/0411111}}.

\bibitem{Vermaseren:2005qc}
J.~A.~M. Vermaseren, A.~Vogt, S.~Moch, {The third-order QCD corrections to
  deep-inelastic scattering by photon exchange}, Nucl. Phys. B724 (2005)
  3--182.
\newblock \href {http://arxiv.org/abs/hep-ph/0504242}
  {\path{arXiv:hep-ph/0504242}}, \href
  {http://dx.doi.org/10.1016/j.nuclphysb.2005.06.020}
  {\path{doi:10.1016/j.nuclphysb.2005.06.020}}.

\bibitem{Baikov:2006ai}
P.~A. Baikov, K.~G. Chetyrkin, {New four loop results in QCD}, Nucl. Phys.
  Proc. Suppl. 160 (2006) 76--79.
\newblock \href {http://dx.doi.org/10.1016/j.nuclphysbps.2006.09.031}
  {\path{doi:10.1016/j.nuclphysbps.2006.09.031}}.

\bibitem{Velizhanin:2011es}
V.~Velizhanin, {Four loop anomalous dimension of the second moment of the
  non-singlet twist-2 operator in QCD}, Nucl.Phys. B860 (2012) 288--294.
\newblock \href {http://arxiv.org/abs/1112.3954} {\path{arXiv:1112.3954}},
  \href {http://dx.doi.org/10.1016/j.nuclphysb.2012.03.006}
  {\path{doi:10.1016/j.nuclphysb.2012.03.006}}.

\bibitem{Velizhanin:2014fua}
V.~Velizhanin, {Four loop anomalous dimension of the third and fourth moments
  of the non-singlet twist-2 operator in QCD} \href
  {http://arxiv.org/abs/1411.1331} {\path{arXiv:1411.1331}}.

\end{thebibliography}
\end{document}